\documentclass{aa} 
\usepackage{graphicx}
\usepackage{color}
\usepackage{txfonts}
\usepackage{amstext}

\begin{document}

   \title{Physical properties and chemical composition of the cores in the California molecular cloud}

%   \subtitle{I. Overviewing the $\kappa$-mechanism}

   \author{Guo-Yin Zhang
          \inst{1,2}\fnmsep\thanks{Just to show the usage
          of the elements in the author field}
          \and
          Jin-Long Xu \inst{1}
          \and 
          A. I. Vasyunin \inst{3,4}
          \and
          D. A. Semenov \inst{5,6}
          \and
          Jun-Jie Wang \inst{1}\fnmsep\thanks{Just to show the usage
          of the elements in the author field}
          \and
          Sami Dib \inst{6,7}
          \and
          Tie Liu \inst{8,9}
          \and
%          \\
          Sheng-Yuan Liu \inst{10}
          \and
           Chuan-Peng Zhang \inst{1,6}
          \and
          Xiao-Lan Liu \inst{1}
          \and
          Ke Wang \inst{11,12}
          \and
          Di Li \inst{1,2,13}
          \and
          Zhong-Zu Wu \inst{14}
          \and 
          Jing-Hua Yuan \inst{1}
          \and
          Da-Lei Li \inst{15}
          \and
          Yang Gao \inst{16}
          }

   \institute{National Astronomical Observatories, Chinese Academy of Sciences, Beijing 100101, China\\
              \email{zgyin@nao.cas.cn};\email{wangjj@nao.cas.cn}
          \and
             University of Chinese Academy of Sciences, Beijing 100049, China
           \and
           Ural Federal University, Ekaterinburg, Russia
           \and
Engineering Research Institute ‘Ventspils International Radio
Astronomy Centre’ of Ventspils University of Applied Sciences,
Inženieru 101, Ventspils LV-3601, Latvia
           \and
           Department of Chemistry, Ludwig Maximilian University, Butenandtstr. 5-13,81377 München, 
           Germany
           \and
           Max Planck Institute for Astronomy, K\"{o}nigstuhl 17, D-69117, Heidelberg, Germany   
           \and
           Niels Bohr International Academy, Niels Bohr Institute, Blegdamsvej 17, DK-2100 Copenhagen \O, Denmark
           \and
           Korea Astronomy and Space Science Institute, 776 Daedeokdaero, Yuseong-gu, Daejeon 34055, Republic of Korea
           \and
           East Asian Observatory, 660 N. A'ohoku Place, Hilo, HI 96720, USA
           \and
           Institute of Astronomy and Astrophysics, Academia Sinica. 11F of Astronomy-Mathematics Building, AS/NTU No.1, Sec. 4, Roosevelt Rd, Taipei 10617, Taiwan 
           \and
           Kavli Institute for Astronomy and Astrophysics, Peking University, 5 Yiheyuan Road, Haidian District, Beijing 100871, China
           \and
European Southern Observatory (ESO) Headquarters,
Karl-Schwarzschild-Str. 2,
85748 Garching bei M\"{u}nchen,
Germany    \and
       CAS Key Laboratory of FAST, NAOC, Chinese Academy of Sciences, Beijing 100101, China     
           \and
           College of Physics, Guizhou University, Guiyang 550025, China
           \and
           Xinjiang Astronomical Observatory, CAS, 150, Science 1-street, Urumqi, Xinjiang 830011, China
           \and
           School of Physics and Astronomy, Sun Yat-Sen University, Zhuhai, 519082, Guangdong, China          
           %  \thanks{The university of heaven temporarily does not
           %          accept e-mails}
             }

   \date{Received xx xx, xxxx; accepted xx xx, xxxx}

% \abstract{}{}{}{}{} 
% 5 {} token are mandatory
 
  \abstract
  % context heading (optional)
  % {} leave it empty if necessary  
   {}
  % aims heading (mandatory)
   {We aim to reveal the physical properties and chemical composition of the cores 
   in the California molecular cloud (CMC),   
   so as to better understand the initial conditions of star formation.} 
  % methods heading (mandatory)
   { We made a high-resolution column density map (18.2") with Herschel data, 
   and extracted a complete sample of the cores in the CMC with the \textsl{fellwalker} algorithm. 
   We performed new single-pointing observations of molecular lines near 90 GHz with the IRAM 30m telescope 
   along the main filament of the CMC.  
   In addition, we also performed a numerical modeling of chemical evolution for the cores 
   under the physical conditions.}
  % results heading (mandatory)
   {
    We extracted 300 cores, 
   of which 33 are protostellar and 267 are starless cores. 
   About 51\% (137 of 267) of the starless cores are prestellar cores. 
   Three cores have the potential to evolve into high-mass stars.   
   The prestellar core mass function (CMF) can be well fit by a log-normal form. The high-mass end of the prestellar CMF shows a power-law form with an index $\alpha=-0.9\pm 0.1$ that is shallower than that of the Galactic field stellar mass function. 
   Combining the mass transformation efficiency ($\varepsilon$) from the prestellar core to 
   the star of $15\pm 1\%$ and the core formation efficiency (CFE) of 5.5\%, we suggest an overall
   star formation efficiency of about 1\% in the CMC. 
   In the single-pointing observations with the IRAM 30m telescope, 
   we find that 6 cores show blue-skewed profile, while 4 cores show red-skewed profile. 
   The molecular line detection rates of $\rm C_{2}H (1-0)$, HCN, $\rm HCO^{+} (1-0)$, and HNC 
   in the protostellar cores are higher than those in the prestellar cores. 
   The detection rates of the $\rm H^{13}CO^{+} (1-0)$, $\rm HN^{13}C (1-0)$, and $\rm N_{2}H^{+} (1-0)$ in the cores are higher than reference positions that are offset from the cores. 
[$\rm {HCO}^{+}$]/[HNC] and [$\rm {HCO}^{+}$]/$\rm [N_{2}H^{+}]$ in protostellar cores 
are higher than those in prestellar cores; this can be used as chemical clocks. 
 The best-fit chemical age of the cores with line observations is $\sim 5\times 10^4$~years. 
    }  
  % conclusions heading (optional), leave it empty if necessary 
   {}

   \keywords{   ISM --
                giant molecular cloud --
                astrochemistry --
                astrophysics
               }

   \maketitle
%
%-------------------------------------------------------------------

\section{Introduction}
The California molecular cloud (CMC) is atypical among the local 
giant molecular clouds (GMCs) \citep{Lada2009,Lada2017}. 
The CMC is in an early state of evolution and has not achieved 
the internal physical conditions to promote more active star formation,  
so it is significantly deficient in its star formation activity compared 
to other well-known clouds of its size and mass, such as Orion A and B, Rosette, 
NGC 2264, and W3 
\citep{Kutner1977,Williams1995,Dahm2008,Rivera2013}. 
The CMC shows a filamentary structure with a distance of 450 $\pm$ 23 pc 
and mass of $\rm \sim 10^5\ M_\odot$ \citep{Lada2009}. 
The hottest part of the CMC is located in its southeast, 
and it might be illuminated by a B star $\rm LkH\alpha 101$ \citep{Barsony1991}. 
Although many investigations have been performed in the past decade by  \citet{Lada2009,Harvey2013,li2014,Kong2015,Lada2017,Broekhoven2018}, 
the physical 
properties and chemical composition of the dense cores 
in the CMC are still not well understood. 

The dense cores in molecular clouds  are birthplaces of stars \citep{Alves2007,Lada2008,Andre2014,Hony2015}. 
There is intense debate in the literature on whether the core mass function (CMF) and the initial mass function (IMF) of the newly formed stars %\LEt{this is commonly used for "initial mass function" - is that what you mean here? You also use this later and introduce it again. Please check and rephrase here, and if it is the same thing, remove the introduction in Sect. 3.1.5})
 possess the same underlying mass distribution. A few studies have suggested that this might be the case in a number of low-mass star-forming regions (e.g., \citealt{Alves2007,Knyves2010}). For the Pipe nebula cloud, \citet{Alves2007} argued that the CMF of the cloud is similar to the IMF of the Trapezium cluster \citep{Muench2002}, with an offset that is suggestive of a core-to-star efficiency, $\epsilon,$ of $\approx 0.3$. Recent CMF determinations challenge this picture. \citet{Motte2018} showed that the CMF of dense cores in the W43 massive star-forming region is much shallower than the Galactic field IMF. Furthermore, a number of recent studies have pointed to significant cluster-to-cluster variations in the parameters that characterize the shape of the IMF in young Galactic and extragalactic clusters \citep{Dib2014,Weisz2015,Dib2017,Schneider2018}. From a theoretical perspective, several works have discussed the time-dependent nature of the CMF and how its shape can be affected by ongoing accretion onto prestellar cores (e.g., \citealt{Dib2010}) and in closely packed systems by core coalescence \citep{Shadmehri2004,Dib2007a}. 

Molecular abundance ratios can be used to estimate the evolutionary stages of the dense cores; these are known as chemical clocks. 
Various pairs of molecules have been proposed over time to serve as chemical clocks that can be used to evaluate the evolutionary stages of the dense cores, including pairs of cyanopolyynes \citep{Stahler1984} or oxygen-bearing to nitrogen-bearing species (e.g., \citealt{Doty2002}). Although it was debated whether abundances of chemical species can be used as direct estimates of the physical age of protostellar clouds (see, e.g., \citealt{Gerin2003}), the chemical age can still be used to estimate the evolutionary status of the objects of interest, and it can be compared with well-studied regions of star formation. For example, the young chemical age of the core can be an indication of its young physical age, if the abundances of chemical species are similar to those in a well-studied object that was proven to be young by other studies. 
The CMC is significantly deficient in its star formation activity, 
which makes it a good example on which to confirm whether $\epsilon$ is 
truly a constant, or if it varies with the environment. 
 Furthermore, when a core is chemically young, the question arises whether a set of molecules can trace its evolution. 
In order to address these questions, we need to examine the CMC in more detail.

It is therefore important to understand why the California GMC shows less prominent star formation 
activity than other more active GMCs. To do this, we have to constrain its gas and 
dust properties, such as density, temperature, kinematics, and chemical composition. 
We therefore identified cores in a Herschel dust continuum map, and 
we employed the IRAM 30m antenna to observe the bright lines 
of CO, N$_2$H$^+$, HCN, HNC, HCO$^+$, and CCH and the CO, HNC, and HCO$^+$ isotopologues 
in the individual cores and other 
star-forming environments.
This combination of molecular tracers allows us to probe gas temperature and
density (CO, N$_2$H$^+$, and HCN/HNC), ionization (HCO$^+$ and N$_2$H$^+$), and the local high-energy
radiation intensity (C$_{2}$H/HCN). The paper is organized as follows: In Sect. 2 we describe 
the observations and data reduction. We present and discuss the results in Sects. 3 and 4, respectively. 
A summary is provided in Sect. 5. 

%--------------------------------------------------------------------
\section{Observations and data reduction}
\subsection{Herschel archival data}     
\begin{figure*}
   \centering
   \includegraphics[width=\hsize]{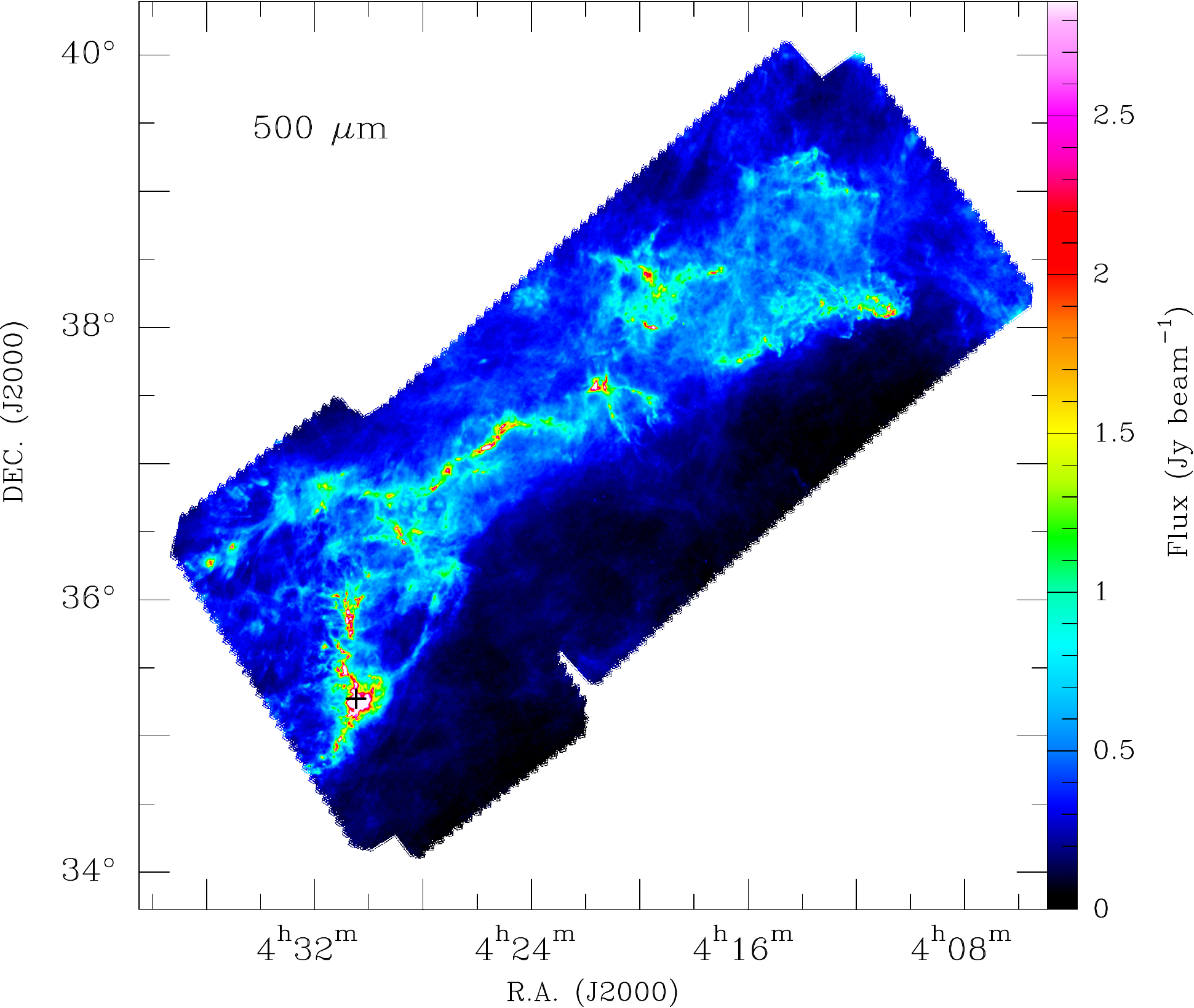}
      \caption{
      California molecular cloud map at the Herschel 500 $\rm \mu m$ band. 
      This map is shown from 0 to 20\% of the peak value (14.3 Jy $\rm beam^{-1}$). 
The 500 $\rm \mu m$ beam is 36.3". 
The area is about 18 $\rm deg^{2}$.
$\rm LkH\alpha 101$ is a B star, which is the brightest star in the CMC. 
It is locates in the lower left corner and illuminates the surrounding gas. 
We mark this star with a black cross on the map. 
}
         \label{Fig:cmcmap}
   \end{figure*}

The Herschel data include PACS 70 and 160 $\rm \mu m$ \citep{Poglitsch2010} and SPIRE 250, 350, and 500 $\rm \mu m$ \citep{Griffin2010} imaging for the CMC \citep{Harvey2013}\footnote{http://irsa.ipac.caltech.edu/data/Herschel/ACMC/}. 
We used the \citet{Harvey2013} version instead of the current HSA pipeline
products, see \citet{Harvey2013}, to determine the details in observations and data reduction processes. 
We focused on a region of the CMC that covers about 18 square degrees.  
An 
image at the Herschel 500 $\rm \mu m$ band for this region is shown in Fig. \ref{Fig:cmcmap}. 
The parallel-mode observations with a fast scan speed $\rm {60}'' s^{-1}$ 
were made with two PACS bands and three SPIRE bands. 
The beam sizes of the PACS at 70 $\mu m$ and 160 $\mu m$ are $\rm {8.4}''$ and $\rm {13.5}''$ , respectively. 
The beam sizes of the SPIRE at 250 $\rm \mu m$, 350 $\rm \mu m$, and 
500 $\rm \mu m$ are $\rm {18.2}''$, $\rm {24.9}''$ , and $\rm {36.3}''$ , respectively.

\subsection{Molecular line observations}
\subsubsection{Source selection}
\begin{figure*}
   \centering
   \includegraphics[width=\hsize]{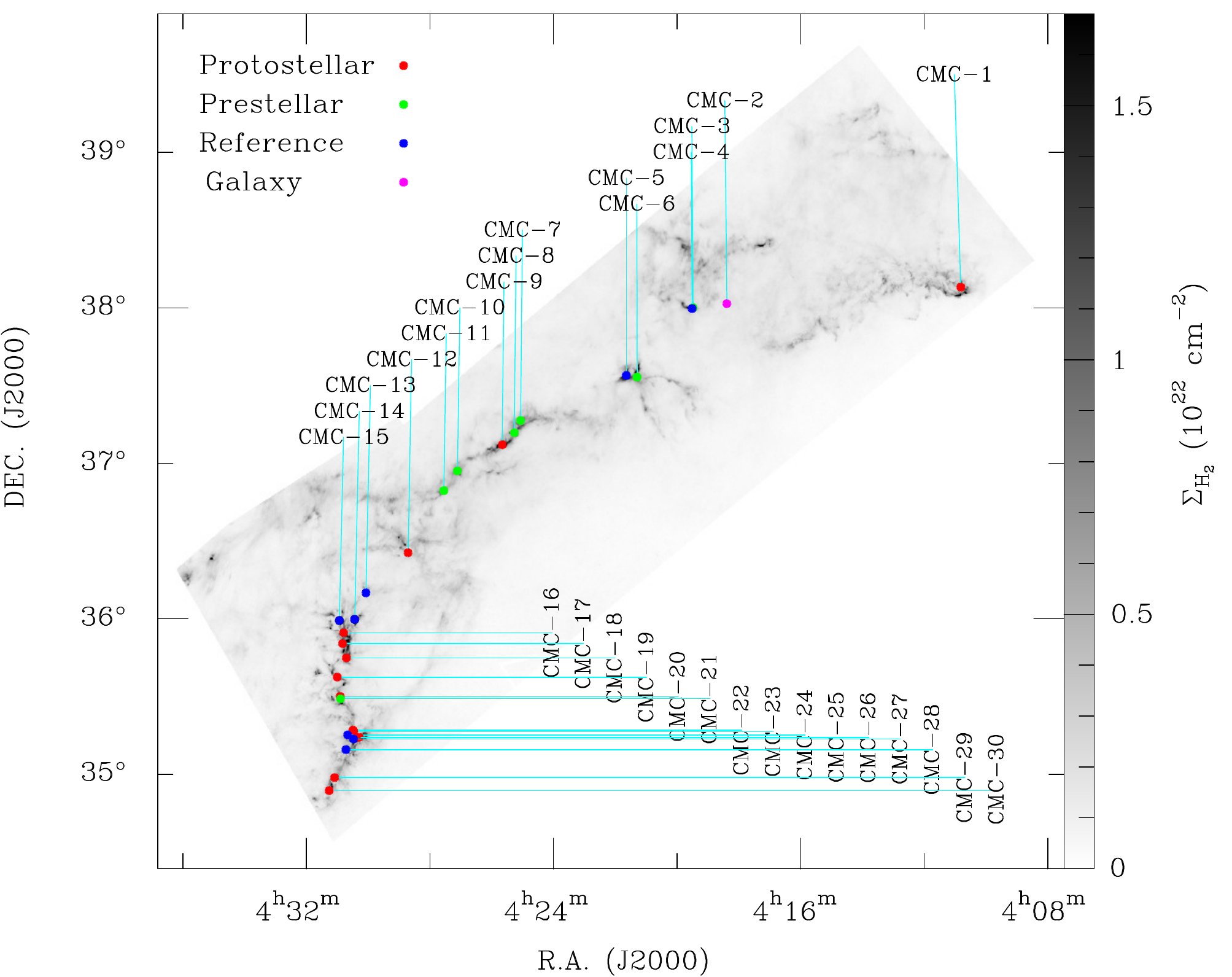}
      \caption{
      Thirty positions for the IRAM 30m single-pointing observations on the  
      Herschel $\rm H_{2}$ column density map. 
      This map is shown from 0 to 20\% of the peak value ($\rm 8.4\times 10^{22}\ cm^{-2}$).  
The 500 $\rm \mu m$ beam is 36.3".
      The 30 observation positions are along the main filament in the CMC. 
The 14 protostellar cores are plotted with red solid circles, 
the seven prestellar cores are plotted with green solid circles, 
the eight observation positions are reference positions that are offset from the cores plotted with blue solid circles, 
and one position is projected on galaxy 3C111, plotted with pink solid circle. 
               }
         \label{Fig:cmcobs}
   \end{figure*}

To explore the chemical properties of the cold dense cores in the CMC, 
we selected 30 positions for single-pointing observations with the IRAM 30m telescope. 
The observed positions are located along the main filament of the CMC. 
Of these 30 positions, 18 are associated with the CSO Bolocam 1.1mm sources \citep{Harvey2013}.  
In order to obtain an unbiased sample with different physical conditions, 
we also selected another 12 high column density positions based on 
the Herschel column density map (\citealt{Harvey2013}). 
These positions are shown in Fig. \ref{Fig:cmcobs} and Table \ref{Tab:obsecmcp}.

\subsubsection{Single-pointing observations with the IRAM 30m}
Single-pointing observations near 90 GHz were carried out in April 2014 using the
IRAM 30m telescope on Pico Veleta, Spain. 
The frequency coverage includes the molecular transitions  
$\rm H^{13}CO^{+} (1-0)$, $ \rm HN^{13}C (1-0)$, $\rm C_{2}H (1-0)$, 
$\rm HCN (1-0)$, $\rm HCO^{+} (1-0)$,  $\rm HNC (1-0)$, 
$\rm N_{2}H^{+}  (1-0)$, $\rm C^{18}O (1-0)$, and $\rm ^{13}CO (1-0)$. 
The single-pixel heterodyne receiver of the Eight MIxer Receiver (EMIR) 
with a bandwith of 16 GHz in two orthogonal polarizations 
was employed to simultaneously observe these nine lines. 
The fast Fourier transform spectrometer (FTS) backends were set to 
200 kHz (about 0.65 km s$^{-1}$ at 90 GHz) resolution. 
We averaged the two orthogonal polarizations.   
We fit linear baselines and subtracted them from the data. 
The molecular lines were converted from the antenna temperature ($\rm T_{A}^{*}$) 
into the main-beam temperature ($\rm T_{mb}$) scale by multiplying with 
the ratio of forward efficiency to main-beam efficiency
\footnote{http://www.iram.es/IRAMES/mainWiki/Iram30mEfficiencies}. 
The forward efficiency from 86 GHz to 115 GHz is from 81\% to 78\%. 
The main-beam efficiency from 86 GHz to 115 GHz is from 95\% to 94\%. 
The beam sizes are ${29}''$ at 86 GHz and ${23}''$ at 110 GHz, 
corresponding to the spatial resolutions of 0.063 pc and 0.05 pc, 
respectively ,at the distance of the CMC (450 pc). 
The typical rms level is 0.05 K for $\rm H^{13}CO^{+} (1-0)$, $\rm HN^{13}C (1-0)$, $\rm C_{2}H (1-0)$, 
$\rm HCN (1-0)$, $\rm HCO^{+} (1-0)$, $\rm HNC (1-0)$, $\rm N_{2}H^{+} (1-0)$, 
and 0.09 K for $\rm C^{18}O (1-0)$ and $\rm ^{13}CO (1-0)$. 
All spectra data were reduced with the 
GILDAS\footnote{http://www.iram.fr/IRAMFR/GILDAS/} software package.
\section{Results}
\subsection{Herschel $H_{2}$ column density results}
\subsubsection{Herschel high-resolution $\rm H_{2}$ column density map}
We applied the \textsl{getsources} tool to derive the column density map of the CMC. 
\textsl{Getsources} \citep{Men2010,Men2012,Men2013} is a  
multi-scale, multi-wavelength source extraction algorithm. 
The bash script \textsl{hirescoldens} in \textsl{getsources} 
was used to create the high-resolution $\rm H_{2}$ column density maps (18.2") 
with Herschel four-band emission from 160 to 500 $\rm \mu m$. The
\textsl{Hirescoldens} methods are described in Appendix A of \citet{Palmeirim2013}. 
The zero-level offsets in the Herschel images are estimated 
from the Planck and IRAS space telescope \citep{Bernard2010}. 
In simple words, this method uses a thin graybody model to fit 
the spectral energy distributions (SEDs) on a pixel-by-pixel basis, 
\begin{equation}
{I}_{v}={B}_{\nu }(T_{d}){\kappa }_{\nu }\Sigma 
 ,\end{equation} 
where ${I}_{v}$ is the surface brightness at frequency $\nu$, 
 $T_{d}$ is dust temperature at each map pixel, 
 ${B}_{\nu }(T_{d})$ is the Planck blackbody function, 
 and we assumed a dust opacity law ${\kappa }_{\nu }=0.1\times (\nu/1000\rm\ GHz )^{\beta }$ $\rm {cm}^{2}g^{-1}$ with $\beta = 2$ \citep{Hildebrand1983,Palmeirim2013}, and
$\Sigma $ is the surface density distribution.
\subsubsection{Core extraction from the $\rm H_{2}$ column density map}
\begin{figure*}
\centering   
\includegraphics[width=\hsize]{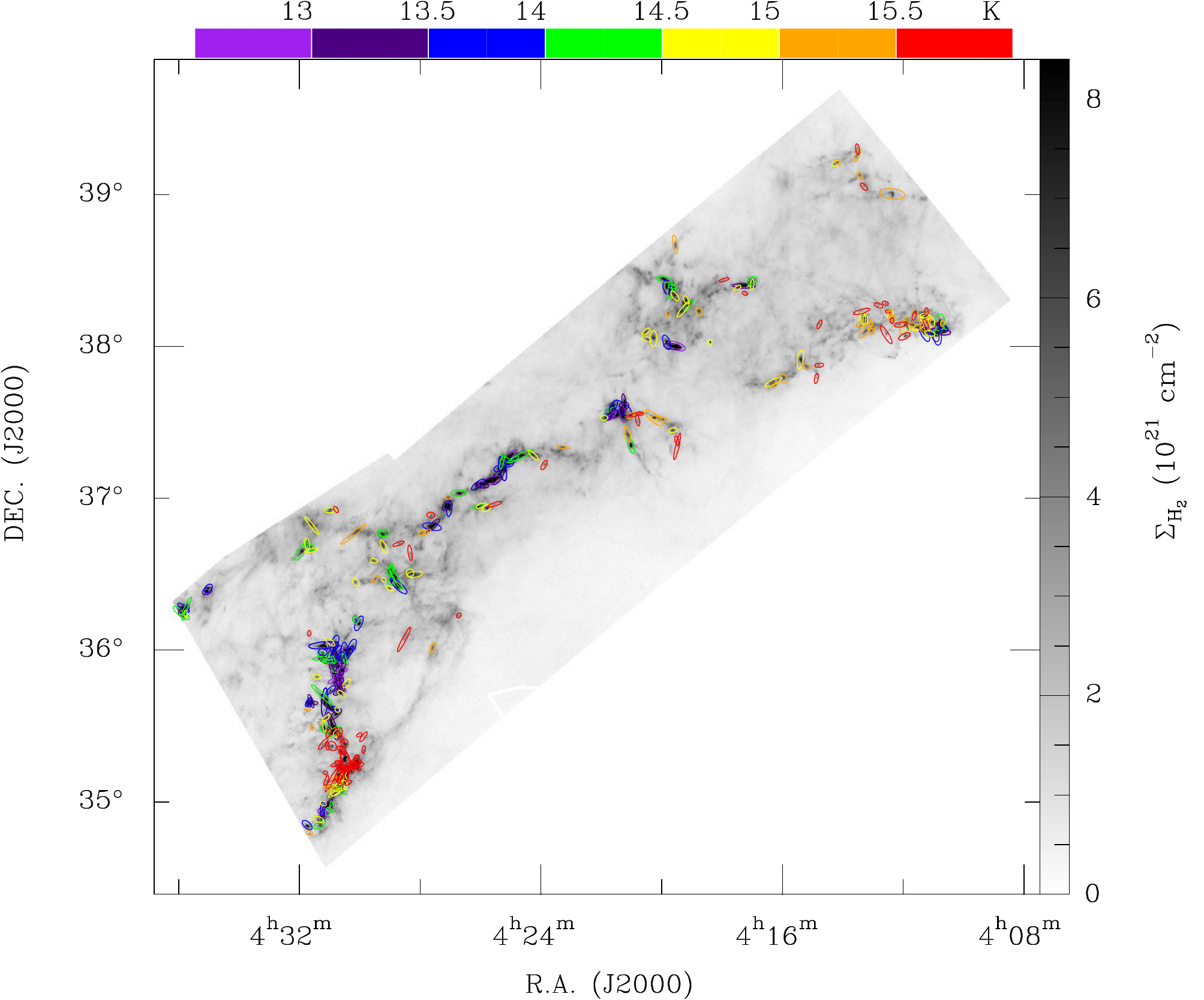}
\caption{ Three hundred cores in the California molecular cloud 
plotted in the $\rm H_2$ column density map. 
This map is shown from 0 to 10\% of the peak value ($\rm 8.4\times 10^{22}\ cm^{-2}$). 
The cores are displayed with the measured dust temperature in units of K, 
coded as shown in the color bar on the top. 
The major and minor axes of the core ellipse shape in this map 
are 4 FWHMs of the equivalent Gaussian. } 
\label{Fig:cmccore}
\end{figure*}

We took two steps to extract dense cores from the column density map.  
First, in order to let the cores stand out, 
we removed the large-scale diffuse structure in the molecular cloud 
with \textsl{cupid findback} \citep{Berry2007}. 
The noise level is $\rm 5\times {10}^{20} {cm}^{-2}$, 
which was determined from off-sources regions in the H$_{2}$ column density map. 
We set the smoothing scale to 95" judged by visual inspection. 
Second, we extracted cores from the column density map using the 
\textsl{fellwalker} algorithm developed by \citet{Berry2015}. 
This algorithm considers all pixels with a value above the noise level. 
It identifies cores by following the steepest gradient route from
each pixel in the map until a significant peak is reached. Each such
peak is associated with a single core, and all pixels along routes
that end at the same peak are assigned to the associated core. Thus
cores extracted by fellwalker are irregular in shape. However,
fellwalker can produce an elliptical approximation to the core shape,
centered on the peak value in the core. It does this by first searching
for the azimuthal angle that gives the largest marginal profile. With
this marginal profile as the major axis, the minor axis is then
assumed to be perpendicular to the major axis. The length of each
marginal profile is the weighted mean of the distances from the peak
to each pixel along the profile, weighted by the pixel data values. 
A scaling factor is then applied that results in the marginal profile
length being equal to the FWHM of the equivalent Gaussian. 
In other words, for a truly Gaussian core, 
the final major and minor axes of the elliptical fit 
would be equal to the FWHM of the Gaussian.  
Finally, we identified 300 cores in the CMC, 
which are shown in Fig. \ref{Fig:cmccore}. 
The detailed \textsl{fellwalker} setup is listed in Table \ref{Tab:FELLWALKER}.
\subsubsection{Physical parameters in the identified cores}
The mass of the core is given by the integral of the column density across the core: 
\begin{equation}
M=\mu _{\rm H_{2}}m_{\rm H}\int {\Sigma}_{\rm {H}_{2}}dA
,\end{equation}
where ${\mu }_{\rm {H}_{2}}=2.8$ is the molecular weight per hydrogen molecule. 
$m_{\rm H}$ is the H atom mass. 
$A$ is the area of the core.  

$L_{\rm maj}$ is the major axis of the ellipse, 
$L_{\rm min}$ is the minor axis. 
They are equal to the FWHMs of the equivalent Gaussian core. 
The effective deconvolved radii of the cores were calculated as 
\begin{equation}
R_{\rm eff}=\frac{\sqrt{\sqrt{(L_{\rm maj}^{2}-{\rm FWHM}_{250\mu m}^{2})}\times \sqrt{(L_{\rm min}^{2}-{\rm FWHM}_{250\mu m}^{2})}}}{2} 
\end{equation}
The volume (V) of the core is $(4/3)\pi R^{3}$ on the assumption that it is spherical. 
Then the number density of hydrogen molecule n(H$_{2}$) in the core 
is $M/(\mu_{\rm H_{2}}*m_{\rm H})/V$.  
The derived masses of the cores are in the range of $\rm 0.1\sim 32.9\ M_\odot$, 
with an average value of 1.9 $\rm M_\odot$. 
The radii of the cores are in the range of $0.01\sim 0.1$ pc, 
 with an average value of 0.04 pc. 
The number densities are in the range of
$\rm 1\times 10^{4}\sim 1.5 \times 10^{6}\ cm^{-3}$,
with an average value of $\rm 1 \times 10^{5}\ cm^{-3}$.
  The dust temperatures and the $\rm H_{2}$ number densities are estimated 
 by averaging the values within the fitting ellipse shape. 
 The dust temperatures of the cores are in the range of 
 $\rm 12 \sim 34$ K, with an average value of 15 K. 
 The core parameters are summarized in Table \ref{Tab:herschelcs}.
\subsubsection{Classification of the dense cores}
\begin{figure*}
\centering   
\includegraphics[width=\hsize]{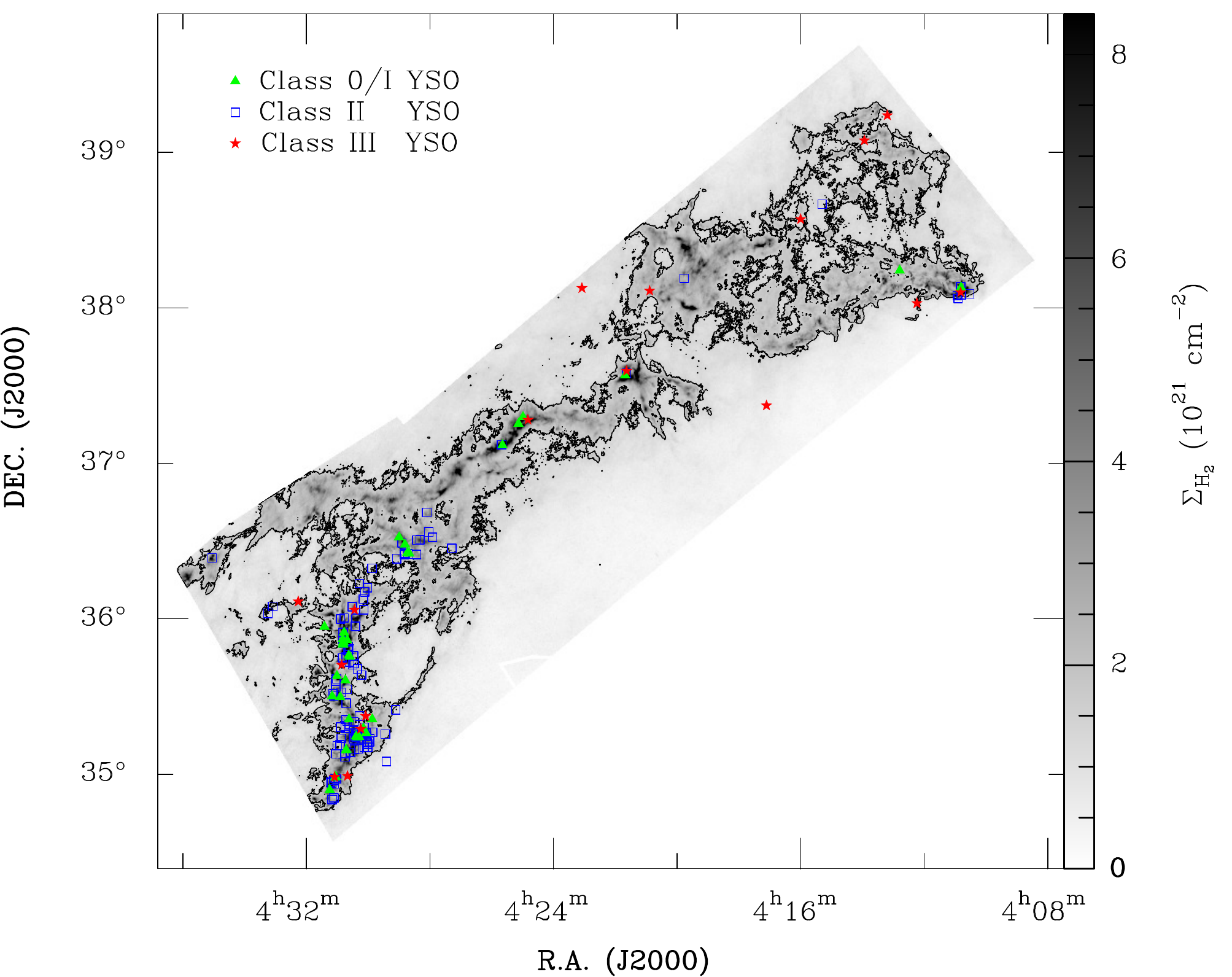}
\caption{Distribution of YSOs in the Herschel $\rm H_{2}$ column density map. 
This map is shown from 0 to 10\% of the peak value ($\rm 8.4\times 10^{22}\ cm^{-2}$).   
The noise $\rm \sigma$ is $\rm 0.5\times {10}^{21}{cm}^{-2}$ estimated in the region from off-sources. 
The black contour is $3 \sigma$.} 
\label{Fig:cmccoldensysos}
\end{figure*}

Based on the presence or absence of young stellar objects (YSOs) detected in the infrared data, 
\citet{Enoch2009} and \citet{Dunham2016} classified the cores into protostellar or starless cores. \citet{Harvey2013} and \citet{Broekhoven2014} have indentified two YSO catalogs in the CMC, respectively. 
In order to resolve the disagreement between these two studies, \citet{Lada2017} reexamined the YSO classifications with the method employed in \citet{Lewis2016}. 
The YSO distribution in the CMC is shown in Fig. \ref{Fig:cmccoldensysos}. 
According to the classification and the new YSO catalog in \citet{Lada2017}, 
we found 33 protostellar cores, accounting for 11\% of the total cores. In addition, 
the prestellar cores are the self-gravitating starless cores. 
The critical Bonnor-Ebert (BE) mass can be used 
to substitute the virial mass \citep{Ebert1955,Bonnor1956,Knyves2010}. 
The critical BE mass can be calculated as 
\begin{equation}
M_{\rm crit}^{\rm BE}\approx \frac{2.4R_{\rm BE}k_{\rm B}T}{G\mu_{\rm p}m_{\rm H} }
,\end{equation}
where $R_{\rm BE}$ is the BE radius, determined by the effective radii of the cores, 
$\mu_{\rm p}$ is the mean molecular weight per free particle. 
Assuming an abundance ratio N(H)/N(He) = 10, $\mu_{\rm p}=2.33$ \citep{Kauffmann2008}. 
Following \citet{Knyves2010}, the assumption of ambient cloud temperature ($T$=10 K) is adopted.  
$G$ is the gravitational constant. 
$k_{B}$ is the Boltzmann constant. 
We identified 137 prestellar cores, accounting for about 51\% of the starless cores.
 \subsubsection{Core mass function and core formation efficiency}
%The initial mass function (IMF) is the mass distribution of the newly formed stars. 
\citet{Salpeter1955} initially proposed a power-law 
IMF:
\begin{equation}
\frac{dN_{\bigstar }}{d{\rm log}M_{\bigstar} }\propto M_{\star }^{-\alpha}, 
{\rm 0.4\ M_\odot}< M_{\bigstar }<{\rm 10\ M_\odot}
,\end{equation}
where $dN_{\bigstar}$ is the number of stars in the mass range $d{\rm log}M_{\bigstar}$ 
and $\alpha =1.35$. 
 \citet{Chabrier2003} postulated a log-normal IMF to replace the spliced power laws,
\begin{equation} 
\begin{matrix}
 &   \frac{dN_{\bigstar }}{d{\rm log}M_{\bigstar }}\propto exp \left [ -\frac{({\rm log}M_{\bigstar }-{\rm log}\mu )^{2})}{2\sigma ^{2}} \right ], M_{\bigstar }\leq {\rm 1\ M_\odot}\\ 
 & \propto M_{\star }^{-1.35}, M_{\bigstar }\geq {\rm 1\ M_\odot}
\end{matrix}
,\end{equation}
where $\mu=0.08$ and $\sigma=0.67$. These parameters changed to $\mu=0.25$ and 
$\sigma=0.55$ in \citet{Chabrier2005}. 
$\mu$ denotes the peak value of the log-normal form IMF 
in units of $\rm M_\odot$ and $\sigma$ is the variance in units of log($\rm M_\odot$). 

\begin{figure*}
\centering   
\includegraphics[width=\hsize]{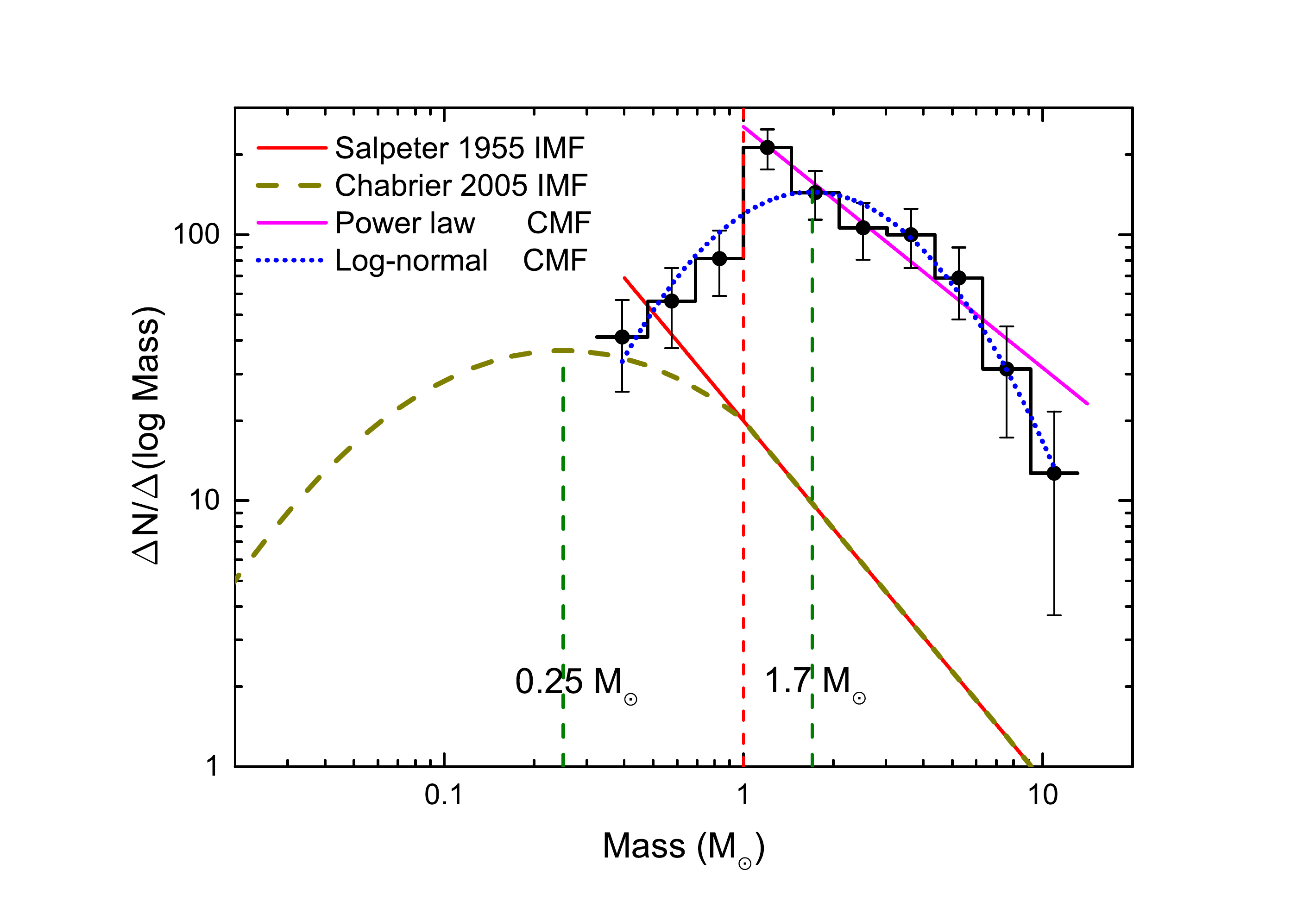}
\caption{Prestellar CMF fit by power law and log-normal distribution. 
 The error bars correspond to $\sqrt{N}$ statistical uncertainties.
 The bin of the prestellar CMF is 0.16 in logarithmic space.
 The number of the bin is 10. 
 The bins start from 1 $\rm {M}_{\odot}$ to the high- and low-mass end. 
 The prestellar CMF observed above 1 $\rm {M}_{\odot}$ can be fit by a power law 
 with $dN/d{\rm log}M\propto M^{-0.9\pm 0.1}$, 
 which is shallower than the \citealt{Salpeter1955} IMF $dN/d{\rm log}M\propto M^{-1.35}$. 
 The prestellar CMF can be well fit by a log-normal with 
 $\rm \mu =log (1.7\pm 0.1\ {M}_{\odot}),\ \sigma =0.37\pm 0.03$ in $\rm log({M}_{\odot})$ unit. 
 1.7 $\rm {M}_{\odot}$ is the log-normal fit peak value, 
 which also corresponds to the log-normal fit average value,
 while $\sigma$ is the standard deviation. 
 The peak value of \citealt{Chabrier2005} is 0.25 $\rm {M}_{\odot}$. 
The prestellar CMF is similar to the IMF.   
 The mass transformation efficiency $\epsilon $ from the prestellar core to the star 
 is about $15\pm 1\%, $ as estimated by comparing the peak value between the CMC log-normal prestellar CMF and the \citet{Chabrier2005} IMF.}
\label{Fig:precmf}
\end{figure*}

The core mass function is the mass distribution of the dense 
%dust 
cores. 
In Table \ref{Tab:herschelcs}, the mass of the identified prestellar 
cores ranges between 0.3 and 13 M$_{\odot}$. 
We fit the prestellar CMF with power-law and log-normal forms. 
Here, we did not take into account the errors in mass estimates,  
but only statistical uncertrainties: $\sqrt{N}$. 
The fitting results are shown in Fig. \ref{Fig:precmf}. 
The prestellar cores heavier than 
1 $\rm M_\odot$ can be well fit with a power-law 
function with $\alpha =-0.9\pm 0.1$, which is shallower than the power-law index 
of \citet{Salpeter1955}. 
The log-normal fitting result is $\mu=1.7\pm 0.1$ and $\sigma=0.37\pm 0.03$. 
By comparing this with the peak value of the stellar mass of 0.25 $\rm M_\odot$ obtained 
from the \citet{Chabrier2005} IMF, we suggest that the mass transformation efficiency 
$\epsilon $ in the CMC from 
the prestellar core to the star is about $15\pm 1\%$. 

In addition, the core formation efficiency (CFE) is the ratio of the total core mass to cloud mass, 
\begin{equation}
CFE=\frac{M_{\rm core}}{M_{\rm cloud}}
.\end{equation} 
We checked the column density map and selected the region without sources to estimate the noise level $\sigma$. $\sigma$ is about $\rm 5\times 10^{20}\ cm^{-2}$. 
The cloud effective mass is $1.1\times 10^4 \rm \ {M}_{\odot}$, 
which was derived by accounting for emission above the 
3$\sigma$ (see Fig. \ref{Fig:cmccoldensysos}). 
The total core mass is 600 $\rm {M}_{\odot}$. Therefore the CFE is 5.5\%. 
\subsection{Molecular-line observations}
\subsubsection{30 single pointings}
The molecular transitions near 90 GHz have typical critical densities ($\rm \gtrapprox {10}^{5}\ {cm}^{-3}$) for collisional excitation, which makes them excellent tracers for probing dense and cold cores \citep{Foster2011}. 
Based on the IRAM 30m telescope, 
30 positions were observed along the main filament in the CMC. 
The properties of molecular lines at the 30 positions are summarized in Table \ref{Tab:mlines}.
We found that 14 positions are protostellar cores,
 7 positions are prestellar cores,  
 8 positions are reference positions that are offset from cores, 
 and 1 position is projected on galaxy 3C111. 

\begin{figure*}
   \centering
   \includegraphics[width=\hsize]{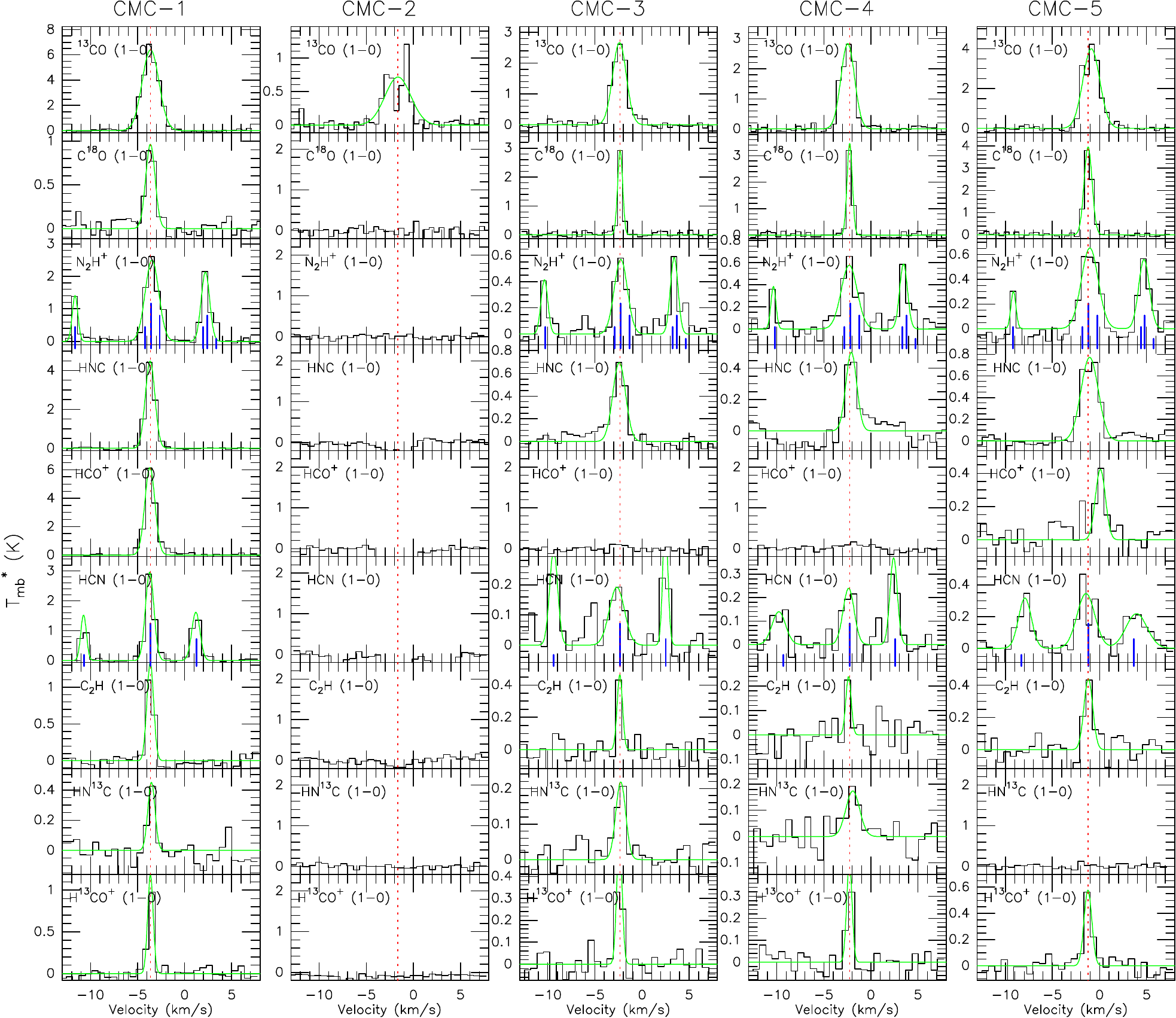}
   \caption{
      Thirty molecular lines of the IRAM 30m observation point. The green curves indicate the Gauss-fit profile. The red dashed lines mark the position of the local standard of rest velocity. The blue lines mark the position of the $\rm N_{2}H^{+}  (1-0)$ and $\rm HCN (1-0)$ hyperfine structure.
              }
   \label{Fig:cmclines}
   \end{figure*}

\begin{figure*}
\centering   
\includegraphics[width=\hsize]{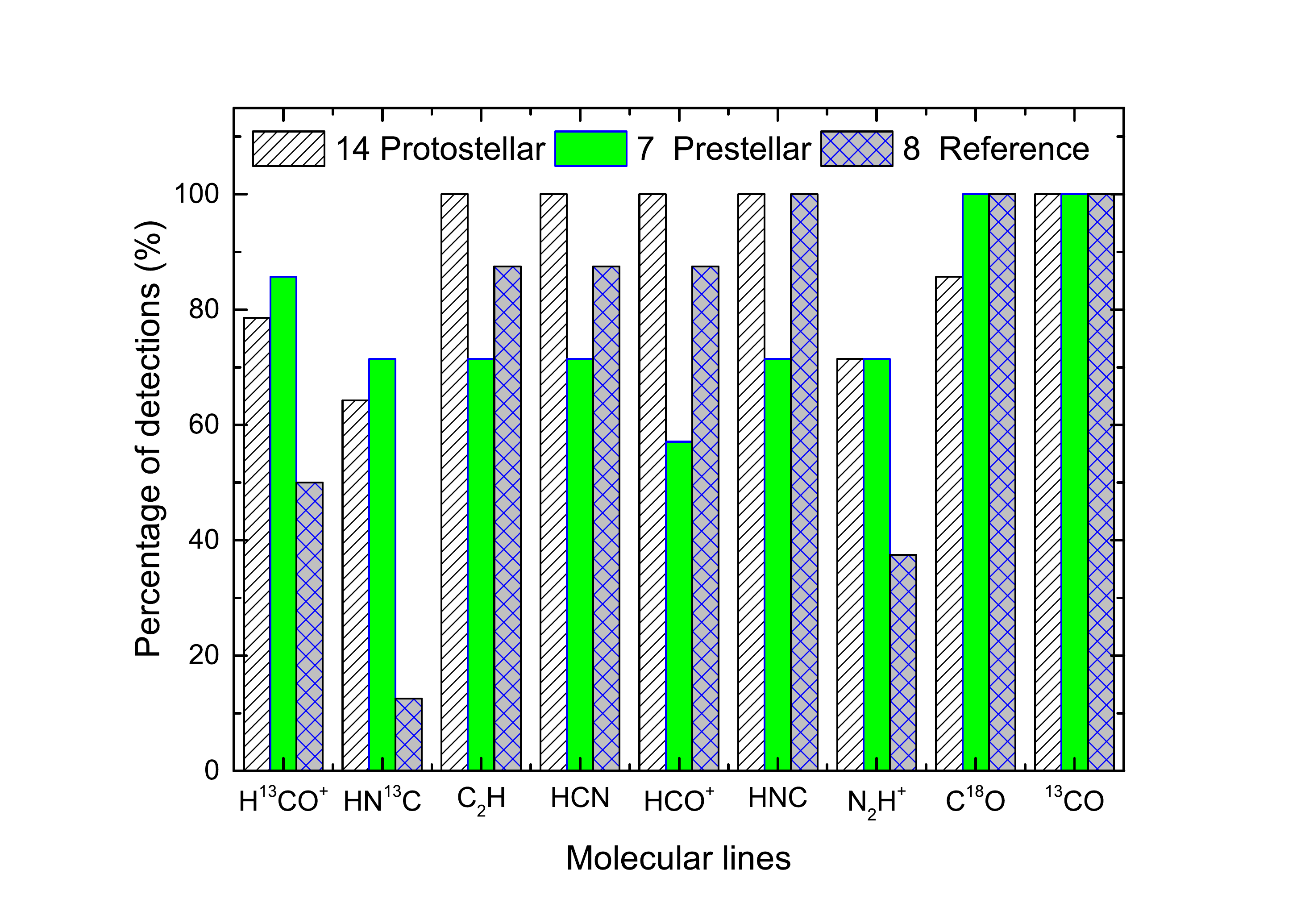}
\caption{Detection rates of the observed molecular lines ($J=1-0$) toward 
14 protostellar cores, 7 prestellar cores, and eight observation positions are reference positions that are offset from cores.} 
\label{Fig:pd}
\end{figure*}

Fig. \ref{Fig:cmclines} shows $\rm H^{13}CO^{+} (1-0)$, $ \rm HN^{13}C (1-0)$, $\rm C_{2}H (1-0)$,
 $\rm HCN (1-0)$, $\rm HCO^{+} (1-0)$, $\rm HNC (1-0)$, 
$\rm N_{2}H^{+}  (1-0)$, $\rm C^{18}O (1-0)$, and  $\rm ^{13}CO (1-0)$ spectra. 
We detected $\rm ^{13}CO (1-0)$ at all the observed positions, 
but $\rm C^{18}O (1-0)$ only at 27 positions.  
Generally, the $\rm C^{18}O (1-0)$ emission is optically thin, 
therefore we used $\rm C^{18}O (1-0)$ to determine the local 
standard of rest velocity ($V_{\rm LSR}$).  For the remaining three position without  
 $\rm C^{18}O (1-0)$ detections, we used $\rm ^{13}CO (1-0)$ to determine the $V_{\rm LSR}$. 
The $V_{\rm LSR}$ 
ranges from -3.63 to 0.52 $\rm km\ s^{-1}$. 
The systemic velocities, line widths, peak intensities, and 
velocity-integrated intensities were obtained by fitting Gaussian profiles. 
The obtained parameters of molecular lines 
at the 30 observed positions are given in Table \ref{Tab:mlines}. 
The detection rates of each molecular line in the 14 protostellar cores, 
7 prestellar cores, and the other reference positions are 
shown in Fig. \ref{Fig:pd}. 
We also give the detection rates of each molecular line in Table \ref{Tab:detecrate}. 
The detection rates of $\rm C_{2}H (1-0)$, HCN, $\rm HCO^{+} (1-0)$, and HNC 
in protostellar cores are higher than that in prestellar cores. 
Other molecular lines are roughly the same. 
We also found that the detection rates of the 
$\rm H^{13}CO^{+} (1-0)$, $\rm HN^{13}C (1-0)$, and $\rm N_{2}H^{+} (1-0)$ 
in protostellar and prestellar cores are higher than that in the reference positions.
\subsubsection{Column density and abundance} 
We used the ratio with the main isotopologue to determine the optical depths for isotopologue pairs 
$\rm H^{13}CO^{+}$ and $\rm HCO^{+}$, and $\rm HN^{13}C$ and HNC. 
Assuming that the line excitation is at local thermodynamical equilibrium (LTE), 
the radiative transfer equation that involves the measured brightness 
temperature ($T_{\rm r}$) is given by 
\begin{equation}
\label{taoe}
T_{\rm r}=[J(T_{\rm ex})-J(T_{\rm bg})]\times [1-exp(-\tau )]f
.\end{equation} 
$T_{\rm mb}$ is the corrected main-beam temperature. 
We have $T_{\rm r}=T_{\rm mb}$.   
$f$ is the beam filling factor. 
$J_{\nu }(T)=\frac{\frac{h\nu }{k}}{{\exp}(\frac{h\nu }{k_{\rm B}T})-1}$. 
$T_{\rm bg}$ is the background temperature of the Universe (2.7 K), 
while $T_{\rm ex}$ is the excitation temperature of the molecular line. 
$\tau$ is the optical depth. 
For the molecular isotopogue pair, $\rm H^{13}CO^{+}$ and $\rm HCO^{+}$, 
we assumed that the former is optically thin, 
but the latter is optically thick. 
We also assumed that they have the same excitation temperature and beam filling factor.  

The optical depths of $\rm H^{13}CO^{+}$ and $\rm HCO^{+}$ were obtained by 
comparing the measured brightness temperatures: 
\begin{equation}
\frac{T_{r}(HCO^{+})}{T_{r}(H^{13}CO^{+})}\approx \frac{1-exp(-\tau _{12})}{1-exp(-\tau _{13})}
.\end{equation}
Furthermore,
\begin{equation}
\frac{\tau _{12}}{\tau _{13}}=\frac{[^{12}C]}{[^{13}C]}
.\end{equation}
The isotopic abundance ratio of $[\rm ^{12}C$/$\rm ^{13}C]$ is in the 
range of $20-70$ \citep{Savage2002}. 
This value is about 20 in the Galactic center, $53\pm 4$ in the 4 kpc molecular ring, 
and $69\pm 6$ in the local interstellar medium (ISM) \citep{Wilson1999}. 
This value in our solar system is 89 \citep{Lang1980}. 
Following \citet{Sanhueza2012}, we assumed a constant 
 abundance ratio of 50 for [$\rm HCO^{+}/H^{13}CO^{+}$] in the CMC.  
We obtained the optical depths of $\rm HN^{13}C$ and HNC with the same approach.  
 
$\rm HCN (1-0)$, $\rm N_{2}H^{+} (1-0)$ and $\rm C_{2}H (1-0)$ have a hyperfine structure. 
We calculated the optical depths of these moleuclar lines 
with the HyperFine Structure (HFS) method in the CLASS software. 
The nuclear spin of the nitrogen nucleus leads to the $\rm HCN (1-0)$ rotational transitions 
showing three components \citep{Loughnane2012}. 
The optical depth for its main component $\rm J F=(12-01)$ was calculated. 
$\rm N_{2}H^{+} (1-0)$ shows seven components \citep{Keto2010}. 
Because closely spaced components overlapped, 
we observed three clusters of lines for $\rm N_{2}H^{+} (1-0)$. 
Two of three lines have three components. 
Only $\rm J F_{1} F=(101-012)$ 
 is separated from other components, 
 through which we estimated the optical depth of this component. 
$\rm C_{2}H (1-0)$ shows six hyperfine components \citep{Padovani2009}. 
We estimate the optical depth of the main component as $\rm JF=(3/2,2-1/2,1)$. 

Under the assumption of LTE, 
the column densities at 30 observation positions were estimated via 
\citep{Mangum2015}
\begin{multline}
N_{\rm tot} = \left(\frac{3h}{8 \pi^3 S \mu^2
  R_{\rm i}}\right)\left(\frac{Q_{\rm rot}}{g_{\rm J} g_{\rm K} g_{\rm I}}\right)
\frac{\exp\left(\frac{E_{\rm u}}{k_{\rm B}T_{\rm ex}}\right)}{\exp\left(\frac{h\nu}{k_{\rm B}T_{\rm ex}}\right)
  - 1}\\
  \times\frac{1}{f\left(J_\nu(T_{\rm ex}) - J_\nu(T_{\rm bg})\right)}\frac{\tau_\nu }{1-\exp(-\tau_\nu )}\int T_{\rm mb} d\upsilon,
  \label{formula1}
\end{multline}
where 
$h$ is the Planck constant.  
$S$ is the line strength. 
$\mu$ is the permanent dipole moment along the axis of symmetry of the molecule 
\footnote{http://www.cv.nrao.edu/php/splat/}. 
$R_{i}$ is the relative intensity normalized to 1.  
$Q_{\rm rot}$ is the rotational partition function. 
$E_{\rm u}$ is the upper energy. 
$g_{\rm J}$ is the rotational degeneracy, 
$g_{\rm K}$ is the K degeneracy, 
and
$g_{\rm I}$ is the nuclear spin degeneracy. 
We assume that $f = 1$. 
For the $J\,\to \,=\,1-0$ transition of linear molecules:
\begin{equation}\label{formula2}
\begin{split}
S & =\frac{J_{\rm u}}{2J_{\rm u}+1}=\frac{1}{3}\\
Q_{\rm rot} & \simeq \frac{\kappa T }{hB_{0}}+\frac{1}{3}\\
g_{\rm J} & =2J_{u}+1=3\\
g_{\rm K} & =1\\
g_{\rm I} & =1
\end{split}
,\end{equation}
where $J_{\rm u}$ is the upper energy level quantum number. 
For $J$ = 1-0, $J_{u}=1$. $B_{0}$ is the rigid rotor rotation constant. 
At high densities of cores ($\rm > 10^{5}\ cm^{-3}$), 
the gas-dust will start to closely couple via collisions (see, e.g., \citealt{Goldsmith1978,Burke1983,Goldsmith2001, Bergin2007}), 
so we adopted $T_{\rm ex}$ as $T_{\rm d}$. 

In contrast, for non-LTE-based calculations of N(X),
multiple transitions of a species X are required,
which we do not have. We have observed the CO isotopologues
HCO$^{+}$, HNC, HCN, $\rm C_{2}H$, and $\rm N_{2}H^{+}$ in their ground rotational
(1-0) transition.  
We examined the validity of the LTE assumption and applied 
this assumption to the column density calculation. 
Usually, in cold sources this occurs when the gas density is higher 
than the so-called critical density by a factor of 10 or more \citep{Pavlyuchenkov2008}. 
The critical density is the gas density that is required to start to populate 
the upper energy level and to excite the line. 
These densities have been computed for various molecules in a number
of papers, see, for example, \citet{Shirley2015b}.  
While for the CO (1-0) line the critical density is $\rm \sim 10^{3}\ cm^{-3}$, 
at low T of $\rm \sim 10$ K.  
In contrast, the critical densities are higher for $\rm HCO^{+} (1-0)$,
$\rm HCN (1-0)$, $\rm HNC (1-0)$ and $\rm N_{2}H^{+} (1-0)$, $\rm > \sim 10^{5}\ cm^{-3}$. This
implies that at $\rm \sim 10$ K, these (1-0) lines could be subthermally excited,
and our column density estimates are underestimated by a factor of several.

\begin{figure*}
\centering   
\includegraphics[width=\hsize]{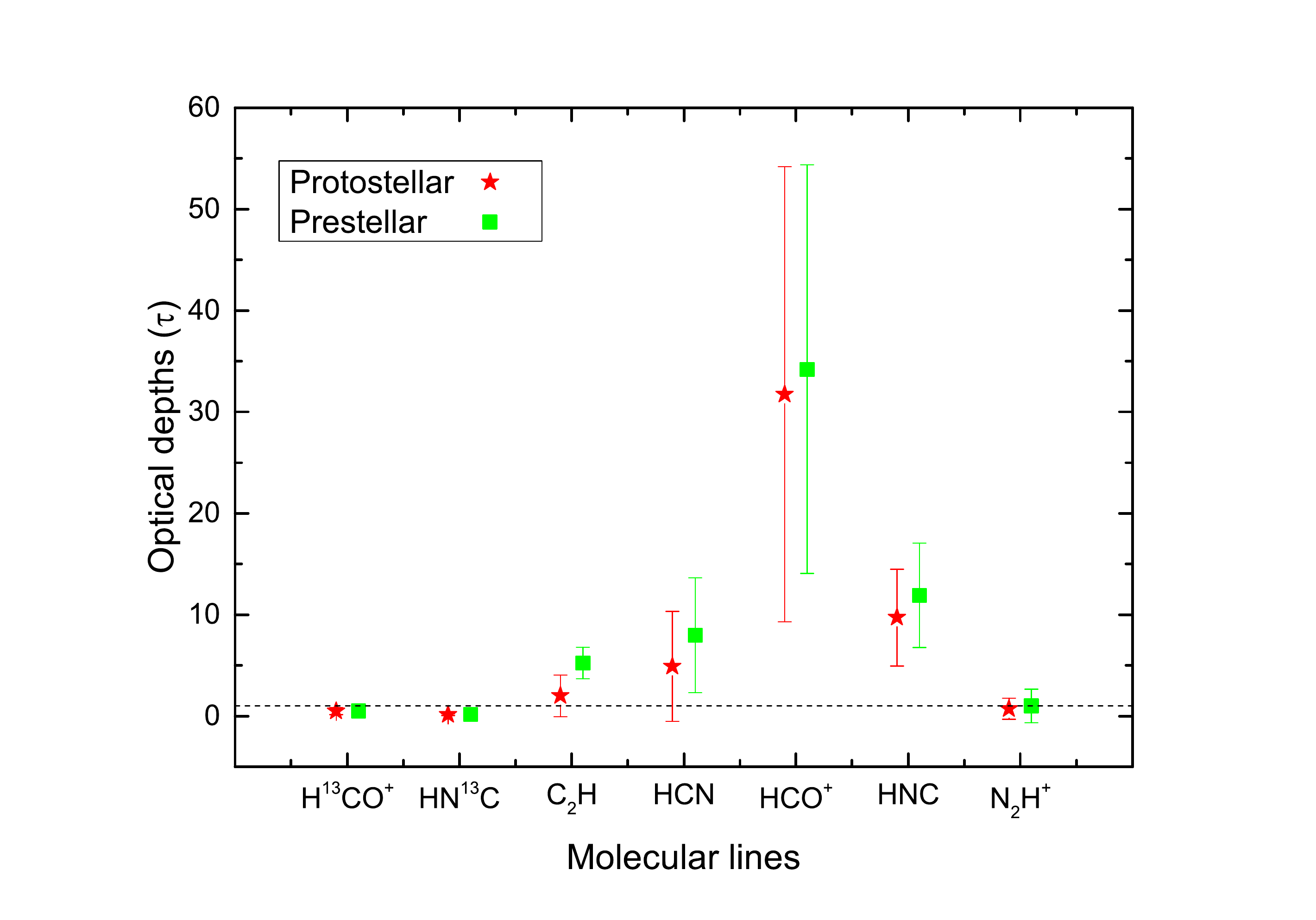}
\caption{Average optical depths ($\tau$ ) of the observed molecular lines ($J=1-0$). The protostellar cores are plotted with red stars and prestellar cores are shown with green squares. The dashed line shows $\tau=1$.} 
\label{Fig:tao}
\end{figure*}

Under the optically thin assumption, the column densities of 
$\rm ^{13}CO$ and $\rm C^{18}O$ are calculated. 
The average optical depths are shown in Fig. \ref{Fig:tao}. 
The calculation parameters are listed in Table \ref{Tab:Paracalcu}. 
The molecular abundances can be defined as $N_{\rm tot}/N(\rm H)$, 
where $N(\rm H)=2\times N(\rm H_{2})$. 
The statistics of molecular abundances 
toward 14 protostellar cores and 7 prestellar cores are given in Table \ref{Tab:statabuncores}.

\subsubsection{Virial masses}
We estimated virial mass from the formula below \citep{MacLaren1988,Bertoldi1992},
\begin{equation}
M_{vir}=\frac{kR\sigma^{2}}{G}
,\end{equation}
where $k$ depends on the core density distribution. 
Following \citet{Parikka2015}, we adopted $k=1.333$, which corresponds to 
a sphere with a power-law density profile $\rho \propto r^{-1.5}$.  
 Here, $\rm C^{18}O (1-0)$ was used to calulate the virial mass. 
The velocity dispersion $\sigma$ is given by
\begin{equation}
\label{velodis}
\sigma =\sqrt{\frac{k_{B}T_{\rm kin}}{\mu m_{\rm H}}+\left ( \frac{\Delta V_{\rm obs}^{2}}{\rm 8ln2}-\frac{k_{B}T_{\rm kin}}{\mu _{\rm tracer}m_{\rm H}} \right )}
,\end{equation} 
where $T_{\rm kin}$ is the kinetic temperature. 
We adopted the dust temperature of core as its kinetic temperature. 
$\mu _{tracer}$ is the mass of the observed molecule. 
$\Delta V_{\rm obs}$ is the observed line width. 
The observed line width consists of thermal and nonthermal components. 
$\frac{kT_{\rm kin}}{\mu _{\rm tracer}m_{\rm H}}$ 
is the thermal component of the observed molecular line, 
while $\frac{\Delta V_{\rm obs}^{2}}{\rm 8ln2}-\frac{kT_{\rm kin}}{\mu _{\rm tracer}m_{\rm H}}$ is 
the nonthermal component. The $\rm C^{18}O(1-0)$ virial masses of our cores are 
in the range of 1.4-5.4 $\rm M_\odot$, with an average value of 3.2 $\rm M_\odot$, 
which is lower than the average mass of the corresponding cores 
obtained from the Herschel column density map. 

\begin{figure*}
\centering   
\includegraphics[width=\hsize]{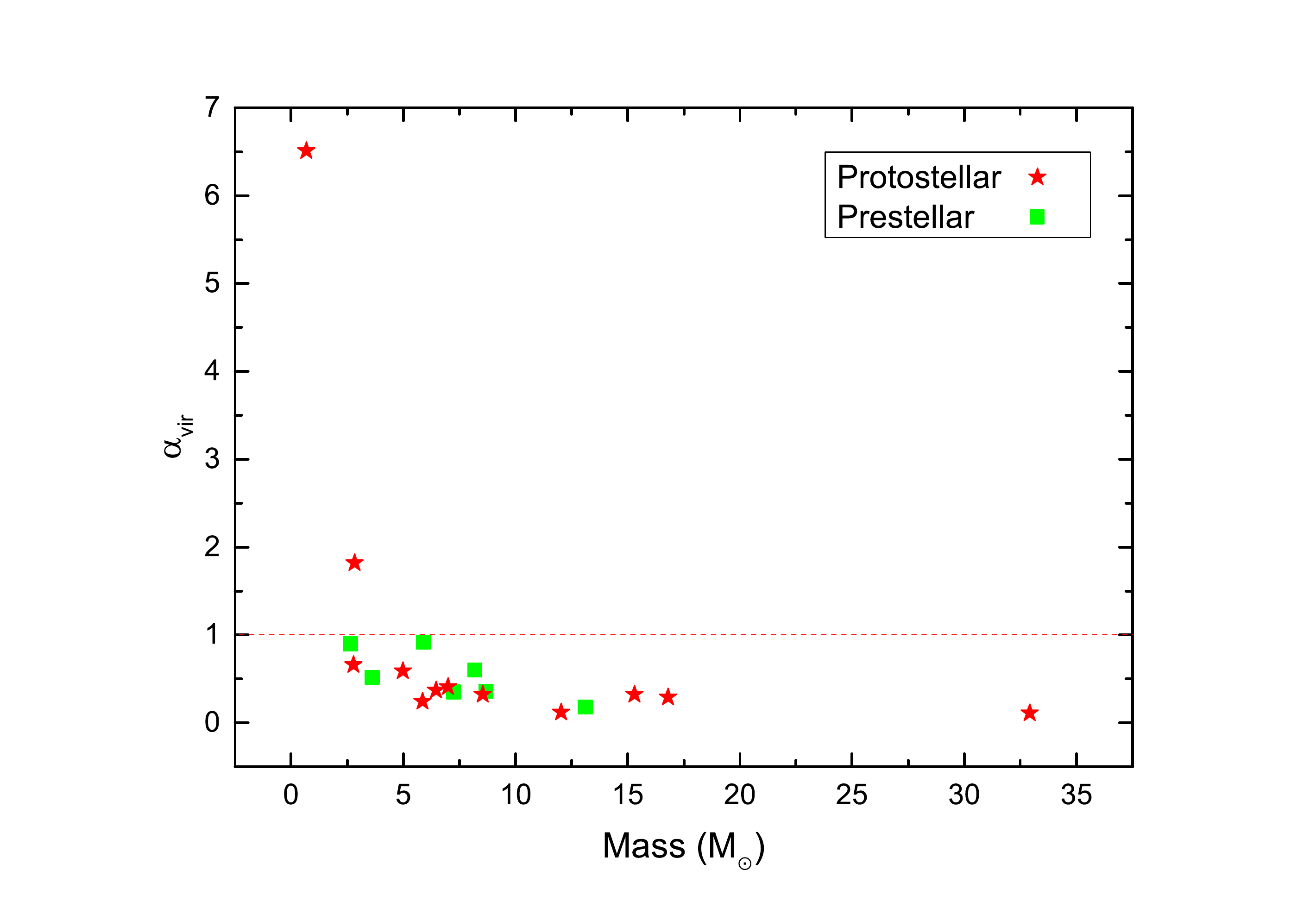}
\caption{$\rm C^{18}O$ virial parameter $\rm \alpha _{vir}$ vs. core masses 
for 12 protostellar cores (2 of 14 protostellar cores do not detect $\rm C^{18}O$) and seven prestellar cores.} 
\label{Fig:virparameter}  
\end{figure*}

The virial parameter can be given by $\alpha _{\rm vir}=M_{\rm vir}/M$, 
where $M$ is the mass of the core. 
%The virial parameter is defined as $\alpha_{vir}=M_{vir}/M$, where $M$ is the physical mass of the core%\LEt{these two sentences appear to say the same, please check and remove one}.
 When we neglect the contributions of the magnetic field and surface terms in the virial equation, it is considered that cores are gravitationally bound whenever $\alpha_{vir}<1$ (e.g., \citealt{Bertoldi1992,Dib2007b}). The virial parameter $\alpha _{\rm vir}$ versus the masses of the cores are displayed in Fig. \ref{Fig:virparameter}. For two prestellar cores, we found that the values of $\alpha_{vir}$ are close to unity, which may indicate that these cores are in a state of near equipartition between gravity and thermal+turbulent support. 
For the $\rm C^{18}O(1-0)$ line, 10 out of 12 of the protostellar cores are gravitationally bound according to their $\alpha_{vir}$ values, and 7 prestellar cores display a similar behavior. This is in agreement with the analysis based on estimates of their BE masses.
\subsubsection{Chemical age}
\begin{figure*}
\centering   
\includegraphics[width=\hsize]{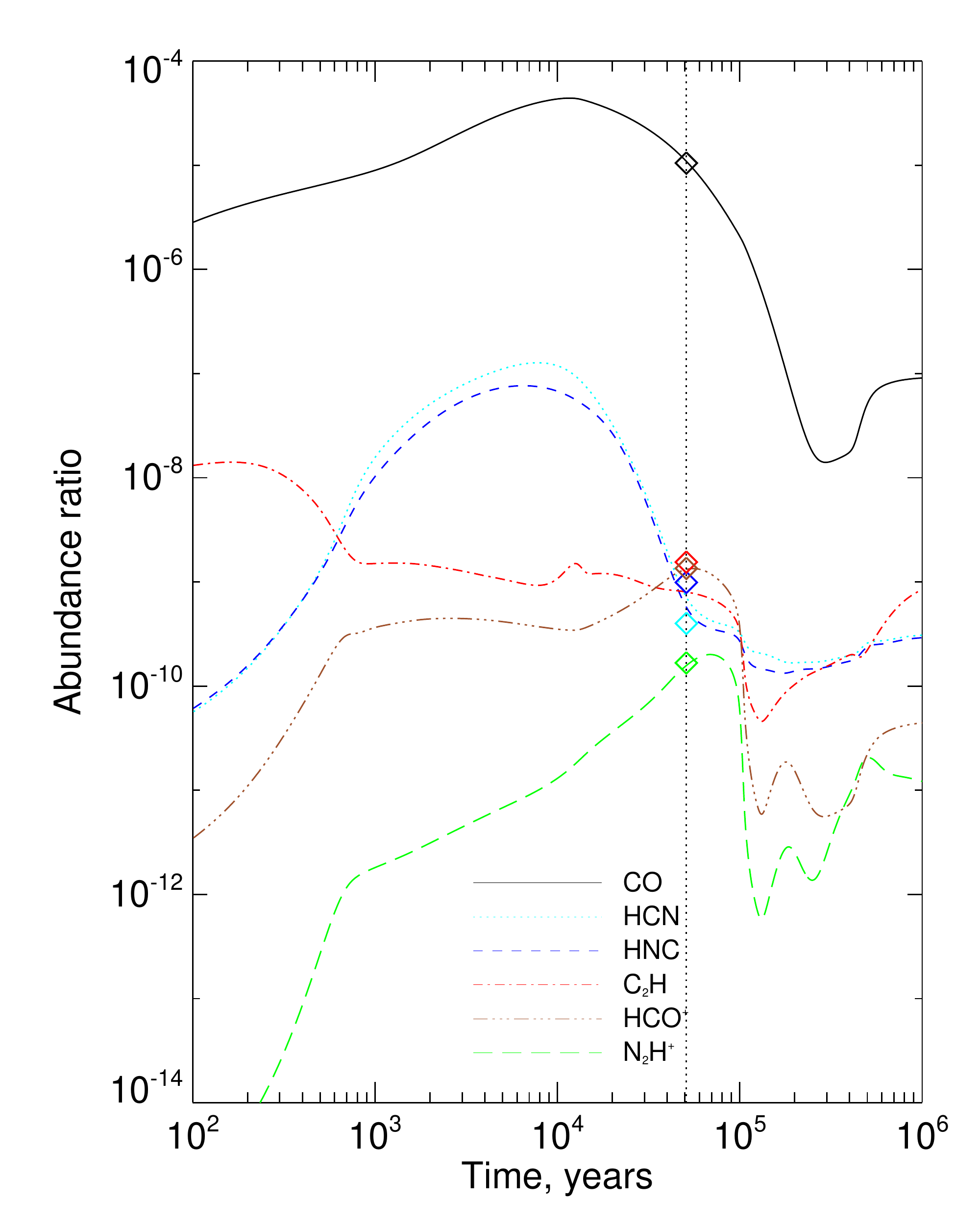}
\caption{ Core chemical age. 
The colored diamonds indicate the average values of the observations from Table \ref{Tab:statabuncores}. 
The CO abundance is obtained from $\rm ^{13}CO$ by [CO]/[$\rm ^{13}CO$]=50. 
The best agreement between the model and the median values of the observations is attained at 
$\sim 5\times 10^4$~years.} 
\label{Fig:chemicalage}  
\end{figure*}

We performed a numerical modeling of chemical evolution under 
the physical conditions typical for a core in the early phase of its 
evolution. We constructed a model of a "typical core" using the information 
on the physical condition from Table \ref{Tab:herschelcs}. 
These are the physical conditions for a "typical core": 
a gas density of $\rm 1\times10^{5}\ cm^{-3}$, 
$T_{\rm gas}=T_{\rm dust}=\rm 15\ $K, and a visual extinction Av = 10. 
The cosmic ray ionization rate and the elemental abundances 
in the CMC are unclear, therefore we took a typical value of $\rm 1.3\times 10^{-17}\ s^{-1}$ 
for the ionization rate, and the typical "low metal" abundances (EA1 in  
\citet{Wakelam2008}) were used as the initial chemical composition. 
We used a 0D numerical gas-grain chemical model, that is, a model without any assumptions about 
the spatial distribution of the species \citep{Vasyunin2009,Vasyunin2013,Vasyunin2017}. 
This chemical model is based on \citep{Vasyunin2017}. 
It employs rate-equation-based three-phase treatment of 
gas-grain chemistry (gas-ice surface - icy bulk) and the chemical network initially 
published in \citet{Semenov2010} and then further modified in \citet{Vasyunin2013} 
and \citet{Vasyunin2017}. In contrast to \citet{Vasyunin2017}, who used a1D model that consisted of  128 0D models and allowed us to calculate 1D distributions of molecules versus radius in the well-studied core L1544, we here calculated just one representative 0D point, because we currently lack knowledge of the inner structure of the cores in CMC 
to basically calculate 
the chemical evolution of a unit volume of gas-dust mixture 
under the provided physical conditions. 
The chemical age is the evolutionary 
time when the modeled abundances of the species match the observed values best. 
We used the modeling results to estimate the 
chemical age of observed cores. 
The best-fit criterion is described in \citet{Vasyunin2017} (Eq. 17). 
The modeling results are shown in Fig. \ref{Fig:chemicalage}. 
There, we show the 
fractional abundances of species versus time. The best agreement
between the model and observations is attained at $\sim 5\times 10^4$~years. 
\section{Discussion}
\subsection{Star formation scenario}

\begin{figure*}
\centering   
\includegraphics[width=\hsize]{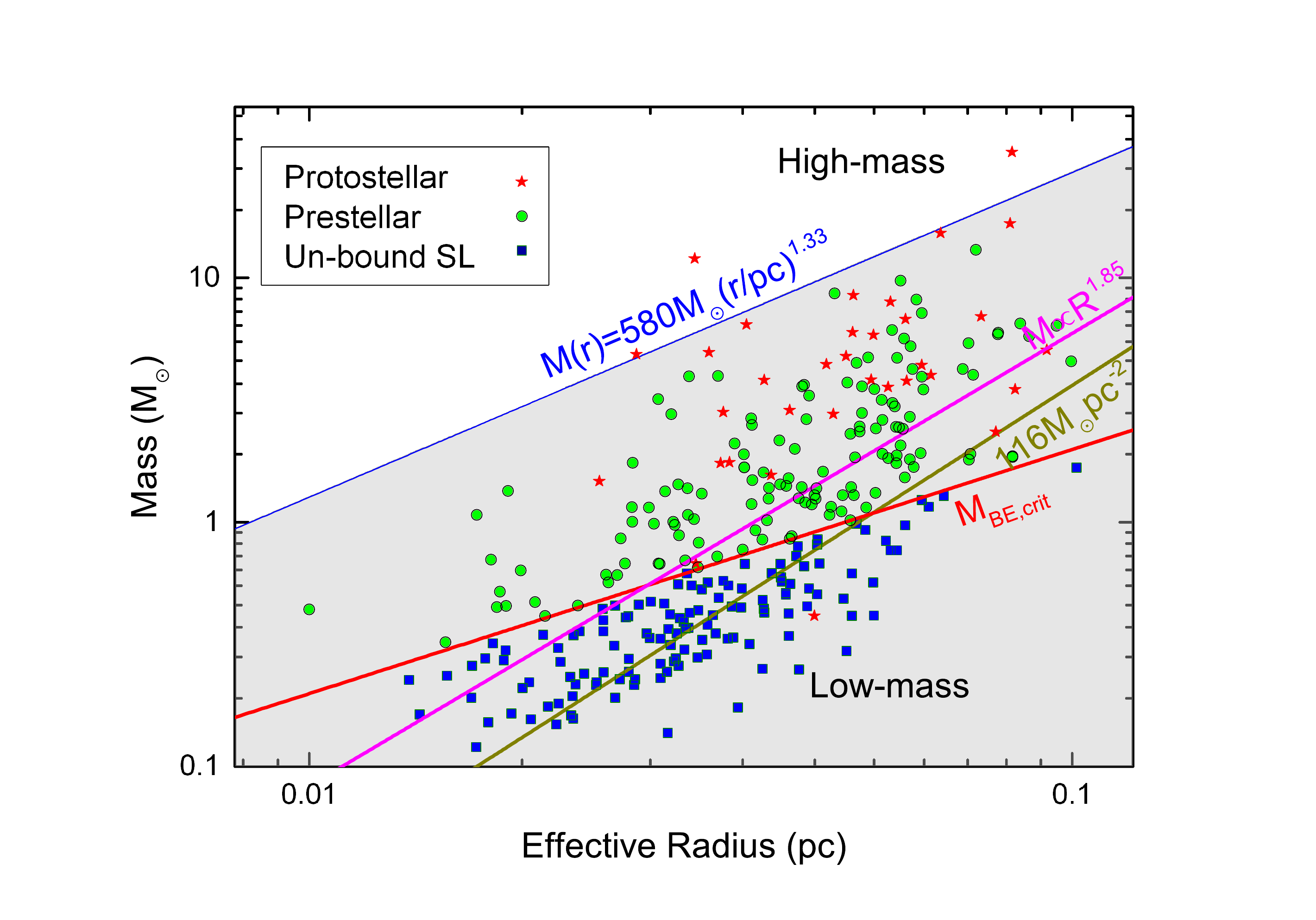}
\caption{Mass-radius relationship of cores. 
The red star, green circle, and blue square indicate protostellar, 
prestellar, and unbound starless cores, respectively. 
The pink line shows that the mass-radius relationship for all the cores can be well 
fit with a power law: $M\propto R^{1.85}$. 
The gray shaded region indicates that high-mass stars cannot form \citep{Kauffmann2010}. 
The green line presents surface densities of 116 $\rm M_\odot\ pc^{2}$. 
The red line presents the surface density of the critical Bonnor-Ebert sphere.}. 
\label{Fig:RM} 
\end{figure*}

From the Herschel high-resolution $\rm H_{2}$ column density map, 
we identified 300 cores in the CMC. According to \citet{Kauffmann2010}, 
if the mass of the core is greater than 
$580{\rm M_\odot} (R/{\rm pc})^{1.33}$, 
where $R$ is the effective radius of the core, 
then they have the potential of forming massive stars. 
Fig. \ref{Fig:RM} presents a mass versus radius plot of the identified cores in the CMC. 
We found that three cores, CMCHerschel-121, CMCHerschel-276, and CMCHerschel-290, lie above the threshold, indicating that they are dense and massive enough to potentially form massive stars. 
The mass of CMCHerschel-121 is about 33 $\rm M_\odot$, while its effective radius is about 
0.08 pc. In our IRAM 30m observations, it is called CMC-9. 
From Fig. \ref{Fig:cmclines}, 
we found that there is a self-absorption dip in the optically thick line 
$\rm HNC (1-0)$ of this core, and a blue-skewed profile for $\rm HCN (1-0)$ and $\rm {HCO}^{+} (1-0)$. 
We therefore suggest that this core is a collapsing high-mass core candidate. 
The mass of CMCHerschel-276 is about 4.9 $\rm M_\odot$; we did not observe it with the IRAM 30m telescope. 
CMCHerschel-290 is called CMC-29 in our IRAM 30m observations. 
The mass is about 12 $\rm M_\odot$ and the radius is about 0.03 pc. 
There are no blue or red profiles in its molecular lines. It is relatively quiet. 

\begin{figure*}
\centering   
\includegraphics[width=\hsize]{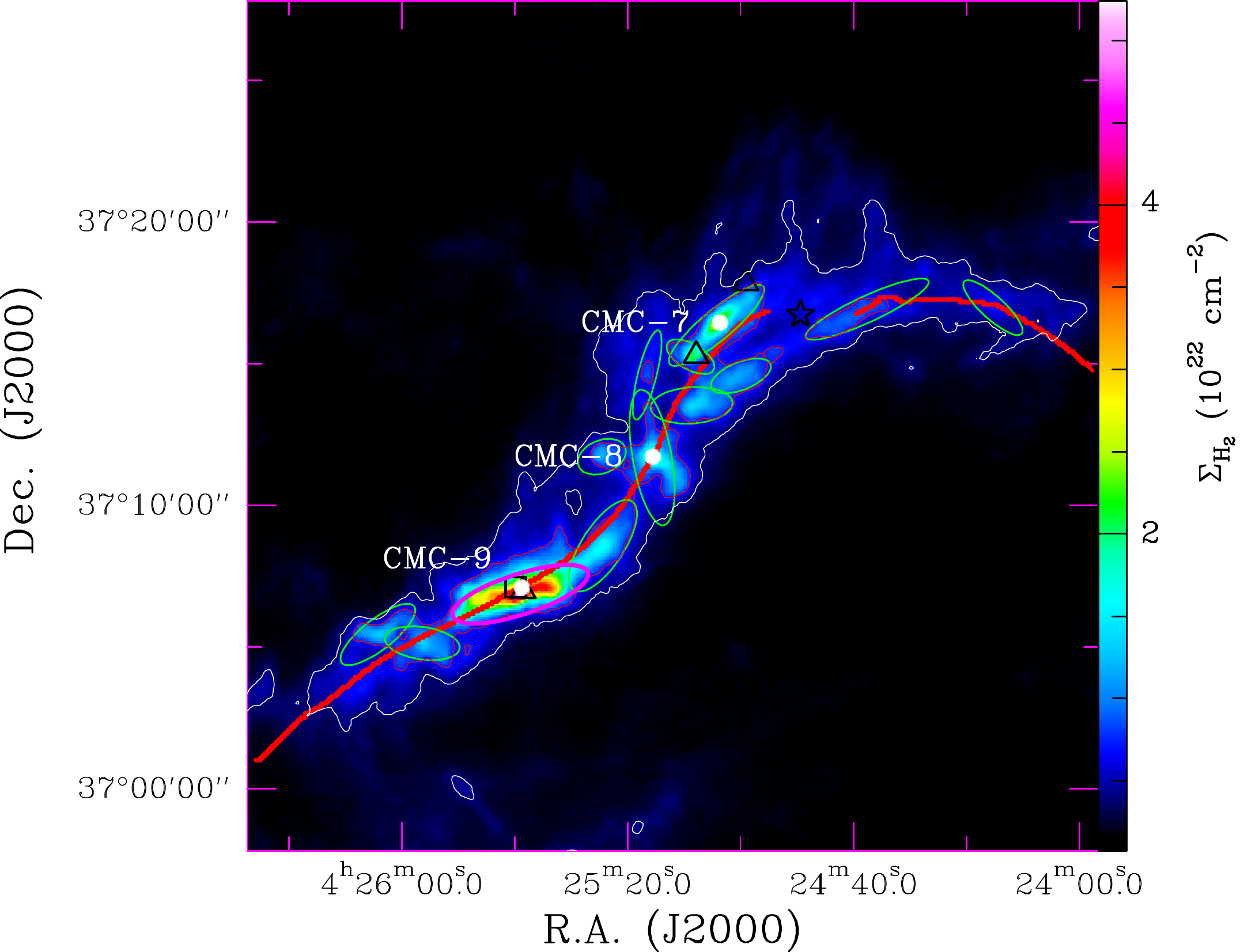}
\caption{Filament existing high-mass core.  
$\rm \sigma=1.3\times {10}^{21}{cm}^{-2}$ is the noise estimated in the local region from off-sources. 
The white contour is $3 \sigma$. 
The pink contour is the average surface density threshold for efficient star formation,
 $\rm 116\ {M}_{\odot}{pc}^{-2}$ \citep{Lada2010}, corresponding to $\rm 7.3\times {10}^{21}{cm}^{-2}$. 
The green ellipses are cores. The high-mass cores are plotted with pink ellipses. 
The white solid circle is the IRAM 30m observation point. 
The red curve is the skeleton of the filament extracted by getfilaments \citep{Men2013}. 
The black square is a class II YSO. The black triangle is a class 0/I YSO. 
The five-pointed star is  a class III YSO.}
\label{Fig:hmc}
\end{figure*}

Fig. \ref{Fig:hmc} shows part of the filament that contains 
the high-mass core (CMCHerschel-121). 
The local noise level $\sigma$ is $\rm 1.3\times 10^{21} cm^{-2}$ in 
the Herschel $\rm H_{2}$ column density map. 
We estimate the filament mass to be above $3\sigma$. 
The mass of this filament is about 291.9 $\rm M_{\odot}$, while the length is 6.6 pc. 
The linear mass is about 44.2 $\rm M_{\odot}\ pc^{-1}$. 
The critical linear mass density for the turbulent pressure to 
dominate the thermal pressure 
can be estimated by 
$(M/l)_{\rm max}=84(\Delta V)^{2}{\rm \ M_{\odot}}\ pc^{-1}$ \citep{Jackson2010}. 
Our IRAM 30m observations include three positions in this filament: CMC-7, CMC-8, and CMC-9. 
The average $\rm C^{18}O (1-0)$ FWHM is 0.84  $\rm km\ s^{-1}$. 
The critical line mass from the $\rm C^{18}O (1-0)$ width is 59.3 $\rm M_{\odot}\ pc^{-1}$, 
suggesting that 
the turbulent pressure can support the filament against gravitational collapse.  

Turbulence plays an important role in star formation processes \citep{Klessen2001,Ballesteros2007,Dib2007b,Liu2012,Federrath2012,Padoan2017}. 
The CMF fit from numerical models of turbulent fragmentation in the molecular clouds 
is more similar to a log-normal form than a power-law form 
(e.g., \citealt{Dib2005,Ballesteros2006,Dib2008,Bailey2013,Gong2015}, but see also \citealt{Padoan2011,Hennebelle2018}). 
We find that the CMF of prestellar cores in the CMC is well fit by a log-normal distribution, 
suggesting that turbulence takes an important role in shaping the CMF. 

Based on the offset between the CMF of the Pipe nebula cloud and the IMF of the Trapezium, \citet{Alves2007} proposed that the core-to-star efficiency, 
$\epsilon$ is about $30\%\pm10\%$.
However, they did not take into account that some of the cores in their sample may 
not be prestellar in nature. Because of the low resolution of the Bolocam 1.1 mm,   
\citet{Enoch2008} only give a lower limit on $\epsilon$ 25\% for 
Perseus, Serpens, and Ophiuchus. 
The prestellar $\epsilon$ is about $40\%$ in Aquila \citep{Andre2010,Konyves2015}. 
\citet{Nutter2007} found a turnover at 1.3 $\rm M_{\odot}$ for SCUBA 850 $\rm \mu m$ 
starless cores in Orion. Compared with the turnover 
at 0.1 $\rm M_{\odot}$ for a \citet{Kroupa2002} IMF, 
the starless $\epsilon$ in Orion is about 8\%. 
Compared with the turnover at 0.25 $\rm M_{\odot}$ for a \citet{Chabrier2005} IMF, 
the starless $\epsilon$ in Orion is about 19\%. 
The prestellar $\epsilon$ of about $15\pm 1\%$ in the CMC is 
lower than that in other molecular clouds. 
The turnover mass of the prestellar CMF in the CMC is $\rm \sim 1.7\ M_\odot$. 
This value is close to the critical BE mass ($1.82\ M_\odot$) for a density 
of $\rm 10^{4} cm^{-3}$ and a gas temperature of 10 K \citep{Lada2008}. 
It is also close to that ($\rm \sim 2-3 M_\odot$), which is an indication 
that thermal fragmentation may be responsible for the CMF in the Pipe nebula \citep{Lada2008}. 
We note that like the CMC, 
the Pipe nebula is a cloud with relatively low star formation for its mass and size. 

The CFE is about 5.5\% in the CMC, which matches the values of 4.9\% and 5.5\% 
in the interarm and spiral-arm regions of the Galactic plane from 
$l=37^{\circ}_{.}83$ to $42^{\circ}_{.}50$ with $\left | b \right |\leq 0.5^{\circ}$ \citep{Eden2013}.
The CFEs for the large-scale structures in our Galaxy are markedly similar in range, 
but this value is greatly affected by the local environment. 
The CFE for W3 GMC is 5\%-13\% in the diffuse region, 
while it is 26\%-37\% in the compressed region 
as a result of the expansion of W 4 H II region. 
The feedback from the H II regions also leads to a higher CFE 
and thus enhanced star formation efficiency \citep{Liu2015,Liu2017,Xu2018}. 
The star formation efficiency (SFE) is the fraction of cloud mass that 
is converted into stars. 
Combining the $\epsilon$ and CFE in the CMC, 
we suggest that the SFE is about 1 \% in the CMC. The typical value in the 
Milky Way is about 2\% \citep{Evans1991}. 

The best-fit chemical age of $\sim 5\times 10^4$~years is 
well consistent with the hypothesis the CMC is in an early state of evolution 
, which is somewhat younger than the chemical age for the prototypical prestellar core L1544
(1.6 $\times$10$^{5}$~years, see, e.g., \citet{Jimnez2016,Vasyunin2017}), 
but there is a difference of one order of magnitude with L134N and TMC-1 
(6$\times$10$^{5}$~years, \citet{Charnley2001}), 
and the value is a factor of $\sim$20 lower than the estimated lifetime of the prestellar cores  
in Aquila (1.2 $\times$10$^{6}$~years) \citep{Konyves2015}. 
Obviously, determining the chemical age with a single-point 0D model is merely a rough estimation, as is the concept of the "chemical age" itself. 
The zero-point in terms of the physical conditions that enter the determination of the chemical age is usually ill defined, and thus, a chemical age is prone to being affected by uncertainty. 
In addition to this, uncertainties in astrochemical models also limit the accuracy of modeled abundances of species to a factor of a few at best for simple species  \citep{Vasyunin2004, Vasyunin2008, Wakelam2005, Wakelam2006, Wakelam2010}. As such, a precise estimation of the chemical age is not physically meaningful. Nevertheless, an order-of-magnitude estimation of this value can be a useful tool to confirm the evolutionary status of protostellar objects 
\citep{Stahler1984,Williams1993,Doty2002,Brunken2014}.

It is also interesting to estimate how stable the best-fit chemical age is to variations in the model parameters, since our definition of a "typical core" as an average of the core parameters from 
Table \ref{Tab:herschelcs} may be considered as somewhat too vague and general. 
We ran a set of models in which the temperatures of gas and dust were varied by $\pm 5$ K
and the gas density was varied by a factor of 2. We found that variations of gas and dust temperature in the range 
$[10-20]$ K shift the best-fit chemical age in the range $[4.0-5.5]\times 10^4$ years, 
while variations in gas density by a factor of 2 vary the chemical age in the range 
$[3.0-7.3]\times 10^4$ years. We can therefore conclude that the chemical age is not a precise value, since it depends on uncertain parameters. 
However, an order-of-magnitude estimation of the chemical age is useful because 
this value in the CMC, which is lower than that in well-studied cores in other clouds, 
implies that the cores in the CMC are indeed young and the CMC is in a very eary evolutionary stage.  
\subsection{Dynamics}
The line widths extracted from single-point observation 
hold important information on the core kinematics. 
The average widths of $\rm ^{13}CO (1-0)$ and $\rm C^{18}O (1-0)$ are 1.6 and 0.92 $\rm km\ s^{-1}$ in the CMC, which roughly corresponds to those in the Planck cold dust clumps (1.3 $\rm km\ s^{-1}$ for 
$\rm ^{13}CO$ and 0.8 $\rm km\ s^{-1}$ for $\rm C^{18}O$) \citep{Wu2012} and those with 
valid detections of $\rm {HCO}^{+}$ or $\rm HCN$ (1.67 $\rm km\ s^{-1}$ for $\rm ^{13}CO$ and 1.08 $\rm km\ s^{-1}$ for $\rm C^{18}O$ ) in \citet{Yuan2016}. 
We calculated the asymmetry parameter \citep{Mardones1997},
$\rm \delta V=({V}_{thick}-{V}_{thin})/d{V}_{thin}$. 
The asymmetry parameter has been used to quantify the asymmetry of an optical thick molecular line, 
where $\rm {V}_{thick}$ is the peak velocity of an optical thick line. 
$\rm {V}_{thin}$ is the peak velocity of an optical thin line. 
$\rm d{V}_{thin}$ is hte full width at half-maximum of the optically thin line. 
A line is considered to have a blue-skewed profile 
if $\rm \delta V < -0.25$, and it is red-skewed if $\rm \delta V > 0.25$. 
The optically thick line $\rm {HCO}^{+} (1-0)$ is a good tracer of the motion of infall. 
Usually, $\rm {HCO}^{+} (1-0)$ and $\rm HNC (1-0)$ are optically thick, 
and $\rm {C}^{18}O(1-0)$ is optically thin. Evidence of infall motion is observable in self-absorbed and optically thick line profiles, which show a combination of a double peak with a brighter blue peak or a skewed single blue peak, or optically thin lines that peak at the self-absorption dip of the optically thick line (\citealt{He2015}).  
We find that six cores show a blue-skewed profile and four cores show a red-skewed profile that 
possibly indicates outflowing  
motions. The skewed profiles of the molecular line are shown in Table \ref{Tab:Asymmetry}.

\subsection{Chemical composition}
\label{relations}
\begin{figure}
\centering   
\includegraphics[width=\hsize]{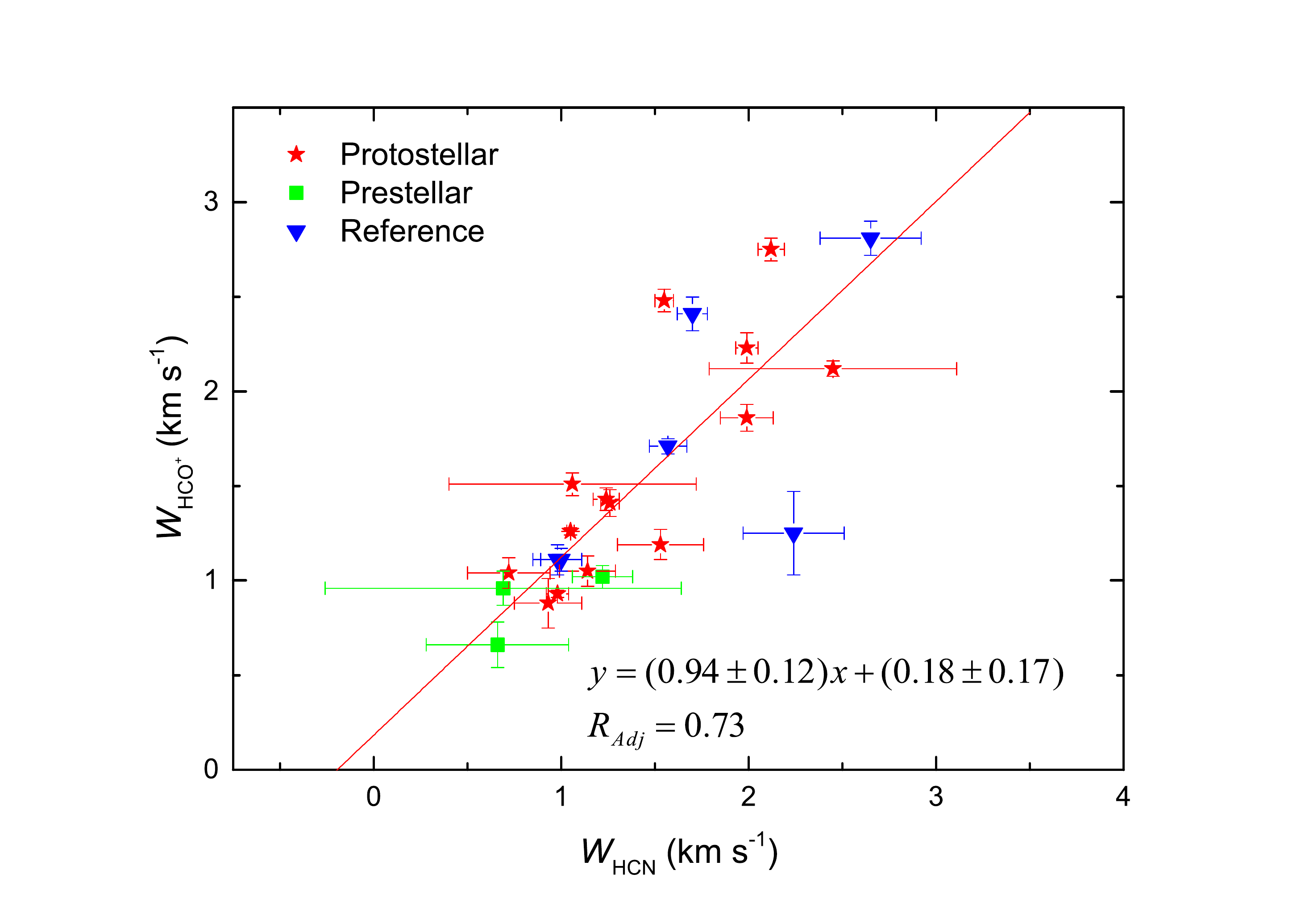}
\includegraphics[width=\hsize]{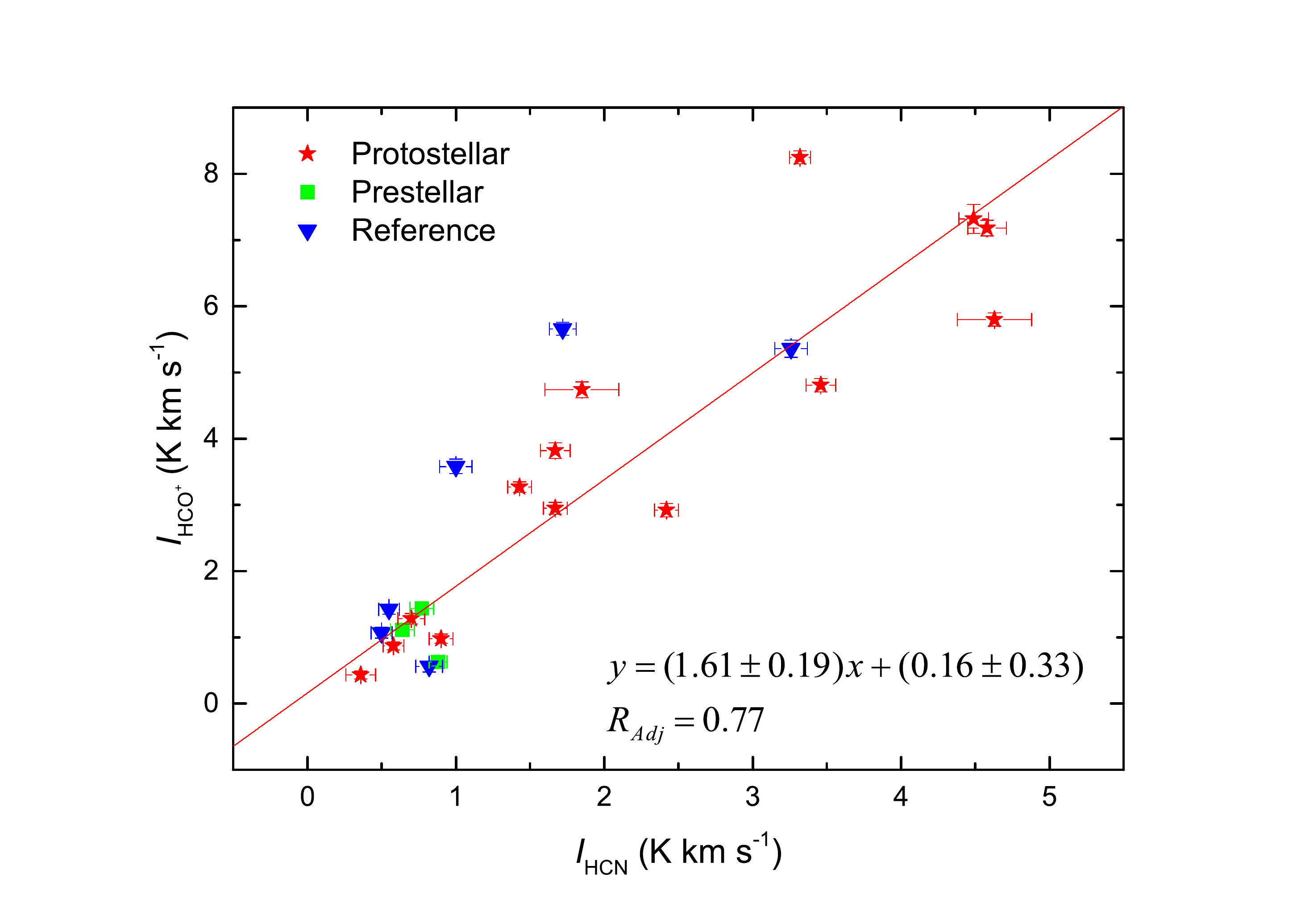}
\caption{Line widths and velocity-integrated intensities 
 of $\rm HCN$ and $\rm {HCO}^{+}$ . They follow strong linear relationships, which indicate 
 that $\rm HCN$ and $\rm {HCO}^{+}$ couple well and there are tight chemical connections.
 The 14 protostellar cores are plotted with red stars, 
 the seven prestellar cores are plotted with green boxes, 
and the eight observation positions are reference positions that are offset from the cores. They are plotted with blue inverted triangles.}
\label{Fig:hcnhcop}
\end{figure}

$\rm HCN$ and $\rm {HCO}^{+}$ have high permanent dipole moments. 
They are good tracers of high-density gas ($\rm \geq 10^{6}\ cm^{-3}$), 
therefore, their emissions are well suited to studying the densest regions in a molecular cloud. 
The line widths of $\rm HCN (1-0)$ and $\rm {HCO}^{+} (1-0)$ at the observed positions are 
in the range of $\rm 0.66\sim 2.65\ km\ s^{-1}$ and 
$\rm 0.66\sim 2.81\ km\ s^{-1}$, 
with an average value of $\rm 1.41\ km\ s^{-1}$ 
and $\rm 1.49\ km\ s^{-1}$ , respectively. 
The velocity-integrated intensities of $\rm HCN$ and $\rm {HCO}^{+}$ are 
in the range of $\rm 0.28\sim 4.63$ and $\rm 0.29\sim 8.25\ K\ km\ s^{-1}$, 
with an average value of $\rm 1.66$ and $\rm 3.04\ K\ km\ s^{-1}$ , respectively. 
We found strong linear correlations between $\rm HCN (1-0)$ 
and $\rm {HCO}^{+} (1-0)$ in line widths and velocity-integrated intensities. 
The line widths are $W_{{\rm HCO}^{+}}=0.94\times W_{\rm HCN}  \text{ and}\ {\rm R_{Adj}}^{2}=0.73$. 
$\rm R_{Adj}^{2}$ is the adjusted R-squared, that is, the fit residuals relative to the error (variance) estimates, which is an important parameter to indicate how well some terms fit a curve or line. 
$\rm R_{Adj}^{2}=0.73$ indicates that about 85\% of the data points can be well 
represented by the fitted line.  
The velocity-integrated intensities are  $I_{{\rm HCO}^{+}}=1.61\times I_{\rm HCN} \text{ and}\ {\rm R}^{2}=0.77$. 
${\rm R}^{2}=0.77$ indicates that about 88\% of the data points can be well 
represented by the fitted line. 
We show these strong linear relationships 
between $\rm HCN$ and $\rm {HCO}^{+}$ in Fig. \ref{Fig:hcnhcop}. 
The intensities of $\rm HCN$ and $\rm {HCO}^{+}$ in infrared dark 
clouds show a linear relation, 
 $I_{{\rm HCO}^{+}}=1.32\times I_{\rm HCN}$ \citep{Liu2013}. 
 \citet{Yuan2016} found that the intensities of $\rm HCN$ and $\rm {HCO}^{+}$ 
 in Planck cold clumps could be fit by a power-law relation, 
 $I_{HCO_{+}}=2.3\times I_{\rm HCN}^{0.8}$. 

The tight correlation between the line widths and line intensities of HCN and HCO$^+$ in
various CMC cores is due to the two-body nature of the chemical processes by which these
species are formed and destroyed, and the lack of strong line excitation gradients in a
typical molecular cloud core \citep{Turner1995,Turner1997}. 
The denser the core, the more readily both of these molecules form through ion-molecule
and neutral-neutral gas-phase reactions. Similarly, the denser the core, the more
efficiently HCO$^+$  dissociates in collisions with electrons, 
while HCN freezes out more efficiently onto dust grains. 

The HCN synthesis is driven by a slow neutral-neutral 
reaction N + CH$_2$ $\rightarrow$ HCN + H, and later, a chemical quasi steady-state is attained. 
This equilibrium is governed by a protonation-recombination cycle: 
HCN + H$_3^+$ $\rightarrow$ HCNH$^+$ +H$_2$, 
HCNH$^+$ + e$^-$ $\rightarrow$ HCN + H, 
with a 34\% probability 
\footnote{http://kida.obs.u-bordeaux1.fr/reaction/2815/HCNH+\_+\_e-.html?filter=Both}. 
The HCO$^+$ chemistry begins with the conversion of almost the entire C budget into CO, which
can then react with H$_3^+$ to form HCO$^+$ and H.  
Later, a chemical steady-state is
attained for HCO$^+$, with the protonation of CO followed by the dissociative recombination
of HCO$^+$ back into H and CO. The direct chemical relationship between HCO$^+$ and HCN through common or 
closely related reactions is quite limited. 
Only one chemical reaction in our chemical model may lead to the 
direct correlation between HCO$^+$ and HCN: HCO$^{+}$~+~HCN $\rightarrow$ HCNH$^{+}$~+~CO. 
However, at least in the model 
considered, this reaction is a destruction channel of moderate importance (up to 15\% of total HCN destruction rate at 10$^5$ years, and up to 7\% of total HCO+ destruction rate at 10$^4$ years of evolution). 

The abundances of the isomer pair HNC and HCN are roughly equal at low temperature 
\citep{Irvine1984,Schilke1992,Ungerechts1997,Graninger2014}. 
The average observed abundances of HNC and HCN are roughly equal in the CMC. 
The value of $\sim 10^{-9}$ for the HNC and HCN abundance also match previous works well 
\citep{Hirota1998,Padovani2011}. 
We found that the average abundance ratio of HCN/HNC in protostellar cores does not have 
a clear difference with that in prestellar cores. 

A combination of the CO optically thin 
isotopologue and N$_2$H$^+$ , for example, can be employed to 
estimate the degree of the CO freeze-out, which can also be used as a coarse indicator
of the core age and the onset of the star formation process.
The main reaction that effectively removes N$_2$H$^+$ from the gas 
when CO is still present is N$_2$H$^+$ + CO $\rightarrow$ HCO$^+$ + N$_2$. Later,
after $\sim 10^5-10^6$~years, when substantial CO freeze-out is expected to occur in dense cores, this
destruction channel becomes inefficient. 
The average abundance ratio of HCO$^+$/N$_2$H$^+$ in protostellar cores is about three times higher than that in prestellar cores. 
We also found that the average abundance ratio of HCO$^+$/HNC in protostellar cores 
is about twice higher than that in prestellar cores.  
There are obvious differences for HCO$^+$/HNC and HCO$^+$/N$_2$H$^+$ in the protostellar cores 
with the prestellar cores. We therefore suggest that HCO$^+$/HNC and HCO$^+$/N$_2$H$^+$ are chemical clocks. 
Our results are similar to those of previous studies (e.g., \citealt{Sanhueza2012,Hoq2013}).

\section{Conclusions}
We made a high-resolution (18.2") $\rm {H}_{2}$ column density map with the Herschel data. 
Using this map, we extracted a complete core sample with the 
\textsl{fellwalker} algorithm.   
In order to estimate the chemical composition and evolution of the cores in the CMC, 
we used the IRAM 30m telescope to carry out new single-point molecular line observations  
near 90 GHz along the CMC main filament. 
These molecular lines include 
$\rm ^{13}CO (1-0)$, $\rm C^{18}O (1-0)$, $\rm N_{2}H^{+} (1-0)$, $\rm HNC (1-0)$, $\rm {HCO}^{+} (1-0)$, 
$\rm HCN (1-0)$, $\rm C_{2}H (1-0)$, $\rm HN^{13}C (1-0)$, and $\rm H^{13}CO^{+} (1-0)$. 
The main results are summarized as follows: 
\begin{enumerate}[(1)]
\item  We extracted 300 cores from the high-resolution $\rm {H}_{2}$ column density map. 
These identified cores contain 33 protostellar cores, 
137 prestellar cores, 
and 130 unbound starless cores. 
The number of the prestellar cores accounts for about 51\% of the starless cores. 
\item The core masses are in the range of $\rm 0.12\sim 33\ M_\odot$ with an average value of 
1.9 $\rm M_\odot$, while the radii are in the range of $0.01\sim 0.1$ pc 
with an average value of 0.04 pc.  
Three cores can evolve into high-mass stars. 
The mass of the high-mass core CMC Herschel-121 is 33 $\rm M_\odot$,  
while its radius is 0.08 pc. 
Moreover, this core, located at a filament, shows an infall feature, 
suggesting that this core may be in a state of collapse. 
 \item 
The prestellar core CMF is slightly steeper than the Galactic field IMF. The prestellar core CMF can be fit a by a lognormal distribution with a peak value $\mu=1.7$ M$_{\odot}$ and $\sigma =0.37\pm0.03$. When fit by a power law in the high-mass end, the exponent of the power law is $-0.9 \pm 0.1$. It should be noted, however, that the CMC is very young, and thus, its population of dense cores is likely to still have ongoing accretion, which will tend to modify or flatten the CMF \citep{Dib2010}. Based on the position of the peak of the CMF with respect to the value of the peak in the \citep{Chabrier2005} IMF, we propose that the core-to-star efficiency, $\epsilon,$ at this stage of its evolution is $15\pm 1\%$. With a CFE of $\approx 5.5 \%$, this translates into an SFE of $\approx 1\%$.
 \item Based on the IRAM 30m telescope single-pointing observations at 30 positions, we find that six cores show a blue-skewed profile and four cores show a red-skewed profile. 
\item The detection rates of $\rm H^{13}CO^{+} (1-0)$, $\rm HN^{13}C (1-0)$, 
and $\rm N_{2}H^{+} (1-0)$ in the protostellar or prestellar cores are clearly higher than 
% than those in than %\LEt{please check this middle part of the sentence again, it is confused and makes no sense in the way it reads now. The rates are higher than the reference positions?} 
the reference positions. 
%are offset from the cores.
Additionally, the detection rates of the molecular lines of $\rm C_{2}H (1-0)$, HCN, $\rm HCO^{+} (1-0)$, and HNC in protostellar cores are higher than those in the prestellar cores.
\item  The line widths and line intensities of $\rm HCN$ correlate well with those of $\rm {HCO}^{+}$. 
[$\rm {HCO}^{+}$]/[HNC] and [$\rm {HCO}^{+}$]/$\rm [N_{2}H^{+}]$ in protostellar cores 
are obviously higher than those in pretellar cores, which can be used as chemical clocks. 
\item  The best-fit chemical age of the cold cores is about $5\times10^{4}$ years, which we derived by fitting the data with an elaborate chemical model. 
\end{enumerate}
 \begin{acknowledgements}
We acknowledge valuable comments from the referee. We thank Charles J. Lada for useful discussion on the manuscript. We thank A. Men?shchikov, D. S. Berry, and S. Bardeau for their technical sup- port with Getsources, Starlink and Gildas, respectively. This work is sup- ported by National Key R\&D Program of China (No. 2017YFA0402600; 2017YFA0402702), National Key Basic Research Program of China (973 Program) (No. 2015CB857101), National Natural Science foundation of China (No. 11703040; 11703074; 11721303; 11763002; U1431111), and Chinese Government Scholarship (No. 201804910583). D.A.S. acknowledges support from the Heidelberg Institute of Theoretical Studies for the project ?Chemical kinetics models and visualization tools: Bridging biology and astronomy?. A.I.V. acknowledges support by the Russian Science Foundation (No. 18-12- 00351). K.W. acknowledges support by the German Research Foundation (grant WA3628-1/1).

\end{acknowledgements}

   \bibliographystyle{aa} % style aa.bst
   \bibliography{bibfile} % your references Yourfile.bib
\begin{table*}
\caption[]{IRAM 30m observed positions.}
\label{Tab:obsecmcp}
 \small
 \begin{tabular}{c c c c c c c c c c c c}
  \hline
  \hline
No. & R.A. & DEC. & $\rm {T}_{dust}$ & $\Sigma_{\rm H_{2}} $ & $R$ & n(H$_{2}$) &$M$ & $M_{V}$ & $M_{BE}$ & Type & Note\\
     & (J2000) & (J2000) & (K) & ($\rm{10}^{21}\ {cm}^{-2}$) & ($\rm pc$)& ($\rm 10^{5}\ cm^{-3}$) & ($\rm M_\odot$) & ($\rm M_\odot$) & ($\rm M_\odot$) & & \\
  \hline\noalign{\smallskip}
   CMC$-$1 &  4h10m41.400s &  38d07m59.00s  &  16.7  &  35.6  &  0.03  &     4.6  &  5.0  &  2.9  &  0.7  &  PRO  &  CMCHerschel-54\\
   CMC$-$2 &  4h18m21.300s &  38d01m36.00s  &  15.1  &  13.2  &  -  &  -  &  -  &  -  &  -  &  Galaxy  &  -\\
   CMC$-$3 &  4h19m27.600s &  38d00m03.00s  &  12.5  &  22.9  &  0.07  &     1.1  &  13.1  &  2.4  &  1.5  &  PRE  &  CMCHerschel-75\\
   CMC$-$4 &  4h19m29.400s &  37d59m43.00s  &  12.4  &  24.6  &  -  &  -  &  -  &  -  &  -  &  REF  &  -\\
   CMC$-$5 &  4h21m38.500s &  37d33m54.00s  &  12.9  &  23.7  &  -  &  -  &  -  &  -  &  -  &  REF  &  -\\
   CMC$-$6 &  4h21m17.400s &  37d33m16.00s  &  12.6  &  21.3  &  0.06  &     1.2  &  8.2  &  4.9  &  1.2  &  PRE  &  CMCHerschel-90\\
   CMC$-$7 &  4h25m03.636s &  37d16m28.83s  &  12.5  &  23.6  &  0.06  &     1.0  &  7.2  &  2.5  &  1.3  &  PRE  &  CMCHerschel-111\\
   CMC$-$8 &  4h25m15.506s &  37d11m44.89s  &  12.8  &  20.2  &  0.08  &     0.4  &  5.9  &  5.4  &  1.6  &  PRE  &  CMCHerschel-119\\
   CMC$-$9 &  4h25m38.700s &  37d07m08.00s  &  12.8  &  39.1  &  0.08  &     2.0  &  32.9  &  3.7  &  1.7  &  PRO  &  CMCHerschel-121\\
  CMC$-$10 &  4h27m04.955s &  36d57m03.17s  &  13.3  &  14.5  &  0.06  &     0.5  &  2.6  &  2.4  &  1.1  &  PRE  &  CMCHerschel-127\\
  CMC$-$11 &  4h27m30.543s &  36d49m26.77s  &  13.2  &  15.5  &  0.05  &     0.8  &  3.6  &  1.9  &  1.1  &  PRE  &  CMCHerschel-135\\
  CMC$-$12 &  4h28m37.500s &  36d25m27.00s  &  13.2  &  21.6  &  0.08  &     0.6  &  7.0  &  2.9  &  1.5  &  PRO  &  CMCHerschel-160\\
  CMC$-$13 &  4h29m56.627s &  36d09m56.60s  &  14.5  &  5.1  &  -  &  -  &  -  &  -  &  -  &  REF  &  -\\
  CMC$-$14 &  4h30m17.258s &  35d59m42.11s  &  13.7  &  7.3  &  -  &  -  &  -  &  -  &  -  &  REF  &  -\\
  CMC$-$15 &  4h30m46.298s &  35d59m16.50s  &  14.0  &  6.7  &  -  &  -  &  -  &  -  &  -  &  REF  &  -\\
  CMC$-$16 &  4h30m37.600s &  35d54m36.00s  &  12.8  &  46.3  &  0.07  &     1.8  &  15.3  &  5.0  &  1.3  &  PRO  &  CMCHerschel-197\\
  CMC$-$17 &  4h30m39.200s &  35d50m22.00s  &  13.1  &  35.4  &  0.04  &     4.3  &  6.5  &  2.4  &  0.7  &  PRO  &  CMCHerschel-205\\
  CMC$-$18 &  4h30m31.300s &  35d44m49.00s  &  12.6  &  17.1  &  0.03  &     2.3  &  2.8  &  5.2  &  0.7  &  PRO  &  CMCHerschel-215\\
  CMC$-$19 &  4h30m47.800s &  35d37m26.00s  &  13.2  &  14.9  &  0.05  &     2.1  &  8.5  &  2.7  &  1.0  &  PRO  &  CMCHerschel-225\\
  CMC$-$20 &  4h30m41.400s &  35d29m58.00s  &  12.8  &  46.3  &  0.08  &     1.0  &  16.8  &  4.8  &  1.7  &  PRO  &  CMCHerschel-232\\
  CMC$-$21 &  4h30m40.800s &  35d29m04.00s  &  13.1  &  34.1  &  0.05  &     2.6  &  8.7  &  3.1  &  1.0  &  PRE  &  CMCHerschel-235\\
  CMC$-$22 &  4h30m16.000s &  35d16m57.00s  &  34.9  &  11.2  &  -  &  -  &  -  &  -  &  -  &  REF  &  -\\
  CMC$-$23 &  4h30m14.385s &  35d16m25.75s  &  34.1  &  9.8  &  -  &  -  &  -  &  -  &  -  &  REF  &  -\\
  CMC$-$24 &  4h30m25.600s &  35d15m07.00s  &  26.2  &  6.2  &  -  &  -  &  -  &  -  &  -  &  REF  &  -\\
  CMC$-$25 &  4h30m03.563s &  35d14m13.88s  &  23.1  &  17.5  &  0.05  &     0.8  &  2.8  &  1.8  &  1.0  &  PRO  &  CMCHerschel-262\\
  CMC$-$26 &  4h30m08.477s &  35d14m01.69s  &  25.0  &  12.1  &  0.03  &     0.7  &  0.7  &  4.4  &  0.6  &  PRO  &  CMCHerschel-264\\
  CMC$-$27 &  4h30m14.397s &  35d13m31.85s  &  24.4  &  9.6  &  -  &  -  &  -  &  -  &  -  &  REF  &  -\\
  CMC$-$28 &  4h30m28.400s &  35d09m27.00s  &  16.2  &  28.1  &  -  &  -  &  -  &  -  &  -  &  REF  &  -\\
  CMC$-$29 &  4h30m48.500s &  34d58m37.00s  &  13.1  &  57.8  &  0.03  &    12.7  &  12.0  &  1.5  &  0.6  &  PRO  &  CMCHerschel-290\\
  CMC$-$30 &  4h30m58.187s &  34d53m36.72s  &  12.8  &  14.5  &  0.05  &     1.2  &  5.9  &  1.4  &  1.1  &  PRO  &  CMCHerschel-296\\
\hline
\end{tabular}
\tablefoot{\\
R.A. and DEC are for the single-pointing observation positions.\\
$\rm {T}_{dust}$ is the dust average temperature in one beam (29", IRAM 30m 86 GHz).\\
$\Sigma_{\rm H_{2}} $ is the $\rm H_{2}$ average column density in one beam.\\
CMC$-$2 is galaxy 3C111. \\
$M$ is the Herschel core mass. $M_{BE}$ is the Herschel core critical Bonnor-Ebert mass. \\
PRE is the gravitationally bound prestellar core. PRO is the protostellar core.\\
REF is the reference position that is offset from the cores.\\
We list the corresponding Herschel core in the notes.
}
\end{table*}

 \begin{table}
\caption[]{FELLWALKER configuration parameter values}
\label{Tab:FELLWALKER}
\setlength{\tabcolsep}{2pt}
\begin{tabular}{r l l}
  \hline
  \hline
Name        & Explanation     & Value                 \\
  \hline\noalign{\smallskip}
RMS                &   estimated at the region without sources   & $\rm 5\times 10^{20}$\\ 
NOISE              &   noise level                               &RMS\\ 
FLATSLOPE          &   change the route gradient                 &RMS\\
MINDIP             &   minimum dip between two adjacent cores    &1.5*RMS\\
MINHEIGHT          &   minimum peak amplitude of a core          &3*RMS\\
FWHMBEAM           &   beam at Herschel 250 $\mu m$, in pixels   &3.64\\
ALLOWEDGE          &   cores are rejected at edge of the data    &0\\
CLEANITER          &   cleans up core edge                       &5\\
MINPIX             &   minimum pixel value in a core             &10\\
\hline
\end{tabular}
\tablefoot{We list the parameters changed from the default values recommended by FELLWALKER}
\end{table} 

\clearpage

\begin{table*}
\caption[]{Parameters of dust dense cores obtained from the Herschel $\rm H_{2}$ column density map.}
\label{Tab:herschelcs}
 \begin{tabular}{c c c c c c c c c c c c}
  \hline
  \hline
No. & R.A. & DEC. &  $L_{\rm maj}$ & $L_{\rm min}$ & $\theta$ & $R_{\rm eff}$  & ${T}_{\rm dust}$& 
n(H$_{2}$) & $M$ & $M_{\rm BE}$ & Type\\
 & (J2000) & (J2000) & (${}''$) & (${}''$)& ($^{\circ}$)& (pc)  &(K)  & ($\rm 10^{5}\ cm^{-3}$) & ($\rm M_\odot$) & ($\rm M_\odot$) & \\
  \hline\noalign{\smallskip}
CMCHerschel$-$1  &  04:13:16.819  &  +39:18:00.73  &    62.7  &    22.3  &    10.0  &    0.03  &  15.6  &     0.3  &     0.3  &     0.6  &  USL  \\
CMCHerschel$-$2  &  04:13:17.652  &  +39:15:32.66  &    80.0  &    26.0  &   156.0  &    0.04  &  15.1  &     0.3  &     0.6  &     0.8  &  USL  \\
CMCHerschel$-$3  &  04:13:58.065  &  +39:12:49.29  &    40.9  &    21.6  &    45.1  &    0.02  &  15.0  &     1.4  &     0.5  &     0.4  &  PRE  \\
CMCHerschel$-$4  &  04:14:01.421  &  +39:12:18.19  &    53.6  &    30.6  &   108.1  &    0.04  &  14.7  &     0.6  &     0.9  &     0.8  &  PRE  \\
CMCHerschel$-$5  &  04:13:13.791  &  +39:07:31.51  &    44.0  &    27.1  &    34.0  &    0.03  &  15.4  &     0.5  &     0.4  &     0.6  &  USL  \\
CMCHerschel$-$6  &  04:13:05.678  &  +39:03:08.46  &    55.8  &    26.0  &    42.0  &    0.03  &  15.8  &     0.3  &     0.4  &     0.7  &  USL  \\
CMCHerschel$-$7  &  04:12:07.603  &  +39:00:17.39  &   150.8  &    60.5  &    82.8  &    0.10  &  15.4  &     0.1  &     1.7  &     2.0  &  USL  \\
CMCHerschel$-$8  &  04:19:30.535  &  +38:40:21.32  &   115.9  &    26.8  &     9.0  &    0.05  &  15.3  &     0.3  &     1.3  &     1.0  &  PRE  \\
CMCHerschel$-$9  &  04:19:53.654  &  +38:26:35.52  &    95.6  &    33.8  &    80.0  &    0.06  &  14.1  &     0.4  &     1.9  &     1.1  &  PRE  \\
CMCHerschel$-$10  &  04:17:52.579  &  +38:26:19.50  &    60.5  &    19.0  &   106.7  &    0.02  &  15.9  &     1.1  &     0.2  &     0.4  &  USL  \\
CMCHerschel$-$11  &  04:16:56.175  &  +38:26:04.70  &    67.9  &    39.5  &   118.6  &    0.05  &  14.4  &     0.2  &     1.0  &     1.0  &  USL  \\
CMCHerschel$-$12  &  04:16:55.783  &  +38:24:31.87  &    59.6  &    20.3  &   175.6  &    0.02  &  14.5  &     1.3  &     0.6  &     0.5  &  PRE  \\
CMCHerschel$-$13  &  04:17:12.094  &  +38:23:58.45  &   136.2  &    28.6  &    88.6  &    0.06  &  13.0  &     1.6  &     9.8  &     1.2  &  PRE  \\
CMCHerschel$-$14  &  04:19:35.760  &  +38:23:39.09  &    51.6  &    21.8  &    85.0  &    0.03  &  14.2  &     0.5  &     0.2  &     0.5  &  USL  \\
CMCHerschel$-$15  &  04:19:47.902  &  +38:23:31.52  &    85.8  &    44.7  &     9.0  &    0.06  &  13.5  &     0.5  &     3.5  &     1.3  &  PRE  \\
CMCHerschel$-$16  &  04:17:25.203  &  +38:23:10.39  &    45.1  &    20.7  &   122.6  &    0.02  &  14.9  &     0.5  &     0.2  &     0.4  &  USL  \\
CMCHerschel$-$17  &  04:19:33.984  &  +38:22:58.00  &    35.3  &    22.1  &    16.0  &    0.02  &  14.3  &     0.7  &     0.2  &     0.4  &  USL  \\
CMCHerschel$-$18  &  04:19:41.895  &  +38:22:13.73  &   107.7  &    23.7  &   173.0  &    0.04  &  14.3  &     0.3  &     0.8  &     0.9  &  USL  \\
CMCHerschel$-$19  &  04:17:10.644  &  +38:20:59.06  &    33.0  &    21.3  &    65.6  &    0.02  &  15.8  &     1.1  &     0.2  &     0.4  &  USL  \\
CMCHerschel$-$20  &  04:19:31.838  &  +38:19:58.05  &    69.1  &    29.2  &    39.0  &    0.04  &  14.7  &     0.2  &     0.3  &     0.8  &  USL  \\
CMCHerschel$-$21  &  04:19:09.104  &  +38:18:20.95  &    52.0  &    26.4  &    19.9  &    0.03  &  14.6  &     0.4  &     0.4  &     0.7  &  USL  \\
CMCHerschel$-$22  &  04:12:28.097  &  +38:17:05.60  &    37.5  &    23.7  &    71.9  &    0.02  &  15.7  &     0.9  &     0.4  &     0.5  &  USL  \\
CMCHerschel$-$23  &  04:19:05.585  &  +38:16:19.73  &    78.6  &    29.5  &   136.9  &    0.05  &  14.2  &     0.5  &     1.3  &     0.9  &  PRE  \\
CMCHerschel$-$24  &  04:12:41.887  &  +38:16:13.29  &    54.8  &    25.5  &    68.9  &    0.03  &  16.0  &     0.3  &     0.3  &     0.7  &  USL  \\
CMCHerschel$-$25  &  04:19:15.209  &  +38:14:33.65  &   115.3  &    26.7  &   140.9  &    0.05  &  14.6  &     0.2  &     0.6  &     1.0  &  USL  \\
CMCHerschel$-$26  &  04:12:22.408  &  +38:13:56.46  &    34.9  &    19.3  &    99.9  &    0.02  &  15.5  &     3.3  &     0.3  &     0.3  &  PRE  \\
CMCHerschel$-$27  &  04:13:16.150  &  +38:13:51.20  &   102.1  &    24.8  &   107.0  &    0.04  &  16.0  &     0.2  &     0.5  &     0.9  &  USL  \\
CMCHerschel$-$28  &  04:11:05.861  &  +38:13:49.44  &    45.4  &    21.5  &   179.7  &    0.02  &  15.5  &     0.6  &     0.2  &     0.5  &  USL  \\
CMCHerschel$-$29  &  04:18:43.352  &  +38:13:46.09  &    51.4  &    36.0  &    14.9  &    0.04  &  15.0  &     0.2  &     0.5  &     0.8  &  USL  \\
CMCHerschel$-$30  &  04:11:28.628  &  +38:13:35.25  &    38.4  &    28.6  &   156.8  &    0.03  &  15.2  &     0.6  &     0.4  &     0.6  &  USL  \\
CMCHerschel$-$31  &  04:12:19.973  &  +38:13:28.24  &    31.7  &    22.5  &     2.9  &    0.02  &  15.4  &     1.4  &     0.3  &     0.4  &  USL  \\
CMCHerschel$-$32  &  04:19:45.621  &  +38:12:46.34  &    36.0  &    18.9  &   168.0  &    0.01  &  15.3  &     2.1  &     0.2  &     0.3  &  USL  \\
CMCHerschel$-$33  &  04:12:18.005  &  +38:12:42.88  &    47.7  &    25.0  &   149.9  &    0.03  &  14.8  &     1.3  &     1.0  &     0.6  &  PRE  \\
CMCHerschel$-$34  &  04:11:10.160  &  +38:12:12.35  &    52.3  &    28.1  &    93.7  &    0.04  &  14.6  &     0.3  &     0.3  &     0.7  &  USL  \\
CMCHerschel$-$35  &  04:11:30.398  &  +38:12:08.82  &    54.4  &    23.1  &   163.8  &    0.03  &  16.0  &     0.3  &     0.2  &     0.6  &  USL  \\
CMCHerschel$-$36  &  04:12:13.672  &  +38:11:22.24  &    44.0  &    23.1  &    41.8  &    0.03  &  15.4  &     0.8  &     0.4  &     0.5  &  USL  \\
CMCHerschel$-$37  &  04:11:08.609  &  +38:11:16.84  &    64.5  &    30.7  &   113.7  &    0.04  &  14.9  &     0.3  &     0.6  &     0.9  &  USL  \\
CMCHerschel$-$38  &  04:13:14.064  &  +38:11:11.98  &    76.2  &    30.4  &     6.0  &    0.05  &  14.8  &     0.3  &     0.8  &     0.9  &  USL  \\
CMCHerschel$-$39  &  04:12:17.450  &  +38:11:06.07  &    31.7  &    27.5  &   167.9  &    0.03  &  15.2  &     1.0  &     0.5  &     0.5  &  USL  \\
CMCHerschel$-$40  &  04:13:09.377  &  +38:10:26.55  &    67.0  &    27.7  &   176.0  &    0.04  &  14.5  &     0.7  &     1.3  &     0.8  &  PRE  \\
CMCHerschel$-$41  &  04:12:15.555  &  +38:10:13.70  &    34.0  &    21.3  &   177.9  &    0.02  &  16.0  &     0.7  &     0.2  &     0.4  &  USL  \\
CMCHerschel$-$42  &  04:11:03.302  &  +38:10:05.31  &    39.7  &    27.7  &   151.7  &    0.03  &  15.2  &     0.5  &     0.4  &     0.6  &  USL  \\
CMCHerschel$-$43  &  04:10:56.385  &  +38:09:34.16  &    63.0  &    39.1  &     4.7  &    0.05  &  14.6  &     0.3  &     1.1  &     1.0  &  PRE  \\
CMCHerschel$-$44  &  04:13:13.095  &  +38:09:13.35  &    90.3  &    32.8  &   110.0  &    0.05  &  15.1  &     0.2  &     0.9  &     1.1  &  USL  \\
CMCHerschel$-$45  &  04:10:34.070  &  +38:09:05.91  &    34.2  &    23.3  &   175.6  &    0.02  &  15.0  &     0.7  &     0.2  &     0.4  &  USL  \\
CMCHerschel$-$46  &  04:11:07.577  &  +38:09:00.21  &    61.0  &    20.7  &    27.7  &    0.03  &  15.6  &     0.5  &     0.3  &     0.5  &  USL  \\
CMCHerschel$-$47  &  04:12:57.365  &  +38:08:48.55  &    29.2  &    21.1  &   154.0  &    0.02  &  15.2  &     2.0  &     0.3  &     0.3  &  USL  \\
CMCHerschel$-$48  &  04:14:41.637  &  +38:08:44.62  &    55.8  &    22.0  &   153.2  &    0.03  &  15.9  &     0.5  &     0.3  &     0.6  &  USL  \\
CMCHerschel$-$49  &  04:10:41.556  &  +38:08:41.03  &   124.4  &    61.4  &   168.6  &    0.09  &  14.4  &     0.2  &     5.1  &     1.8  &  PRO  \\
CMCHerschel$-$50  &  04:11:58.409  &  +38:08:38.90  &    78.9  &    28.3  &    98.8  &    0.04  &  15.8  &     0.3  &     0.7  &     0.9  &  USL  \\
  \hline
\end{tabular}
\tablefoot{\\
R.A. and DEC are the center positions of the core ellipse shape. \\
$L_{max}$ is the major axis of the ellipse, 
$L_{min}$ is the minor axis. 
Their values are equal to the FWHMs of the equivalent Gaussian.\\ 
$\theta $ is the position angle.\\ 
${T}_{\rm dust}$ is the average dust temperature in the ellipse shape.\\
$R_{\rm eff}$ is the core radius deconvolved to remove the effect of the telescope beam.\\
$M$ is the core mass. $M_{BE}$ is the critical Bonnor-Ebert mass. \\
USL is the gravitationally unbound starless core. 
PRE is the bound prestellar core. 
PRO is the protostellar core.
}
\end{table*}

\clearpage

\begin{table*}
\begin{tabular}{c c c c c c c c c c c c}
  \hline
  \hline
No. & R.A. & DEC. &  $L_{\rm maj}$ & $L_{\rm min}$ & $\theta$ & $R_{\rm eff}$  & ${T}_{\rm dust}$& 
n(H$_{2}$) & $M$ & $M_{\rm BE}$ & Type\\
 & (J2000) & (J2000) & (${}''$) & (${}''$)& ($^{\circ}$)& (pc)  &(K)  & ($\rm 10^{5}\ cm^{-3}$) & ($\rm M_\odot$) & ($\rm M_\odot$) & \\
  \hline\noalign{\smallskip}
CMCHerschel$-$51  &  04:11:15.986  &  +38:08:32.97  &    87.6  &    18.6  &    18.7  &    0.02  &  15.2  &     2.1  &     0.5  &     0.4  &  PRE  \\
CMCHerschel$-$52  &  04:11:34.688  &  +38:08:30.77  &   111.1  &    21.7  &    64.7  &    0.04  &  15.1  &     0.5  &     0.9  &     0.8  &  PRE  \\
CMCHerschel$-$53  &  04:11:50.558  &  +38:08:10.21  &   108.3  &    46.3  &   161.8  &    0.07  &  15.1  &     0.2  &     1.9  &     1.5  &  PRE  \\
CMCHerschel$-$54  &  04:10:41.985  &  +38:07:58.98  &    58.5  &    24.9  &    68.6  &    0.03  &  16.4  &     4.6  &     5.0  &     0.7  &  PRO  \\
CMCHerschel$-$55  &  04:10:28.850  &  +38:07:47.28  &    38.8  &    24.3  &    44.6  &    0.03  &  14.0  &     1.8  &     0.9  &     0.5  &  PRE  \\
CMCHerschel$-$56  &  04:11:23.156  &  +38:07:39.04  &    73.5  &    30.5  &   115.7  &    0.05  &  14.6  &     0.4  &     1.2  &     0.9  &  PRE  \\
\hline
\end{tabular}
\centerline{Table \ref{Tab:herschelcs} --- Continued}
\end{table*}

\clearpage

\begin{table*}
 \begin{tabular}{c c c c c c c c c c c c}
  \hline
  \hline
No. & R.A. & DEC. &  $L_{\rm maj}$ & $L_{\rm min}$ & $\theta$ & $R_{\rm eff}$  & ${T}_{\rm dust}$& 
n(H$_{2}$) & $M$ & $M_{\rm BE}$ & Type\\
 & (J2000) & (J2000) & (${}''$) & (${}''$)& ($^{\circ}$)& (pc)  &(K)  & ($\rm 10^{5}\ cm^{-3}$) & ($\rm M_\odot$) & ($\rm M_\odot$) & \\
  \hline\noalign{\smallskip}
CMCHerschel$-$106  &  04:21:01.029  &  +37:20:51.54  &   101.6  &    34.7  &    19.2  &    0.06  &  14.1  &     0.4  &     2.5  &     1.2  &  PRE  \\
CMCHerschel$-$107  &  04:19:28.867  &  +37:20:32.64  &   160.6  &    20.8  &   167.0  &    0.04  &  16.9  &     0.1  &     0.3  &     0.9  &  USL  \\
CMCHerschel$-$108  &  04:23:15.873  &  +37:20:11.19  &    82.1  &    18.5  &    79.5  &    0.02  &  15.3  &     1.6  &     0.3  &     0.4  &  USL  \\
CMCHerschel$-$109  &  04:24:15.242  &  +37:16:59.74  &    82.0  &    24.5  &    47.7  &    0.04  &  14.8  &     0.2  &     0.4  &     0.8  &  USL  \\
CMCHerschel$-$110  &  04:24:37.553  &  +37:16:57.14  &   143.2  &    27.8  &   114.8  &    0.06  &  14.2  &     0.3  &     2.1  &     1.2  &  PRE  \\
CMCHerschel$-$111  &  04:25:04.241  &  +37:16:23.13  &   128.8  &    32.3  &   129.8  &    0.06  &  12.8  &     1.0  &     7.2  &     1.3  &  PRE  \\
\hline
\end{tabular}
\centerline{Table \ref{Tab:herschelcs} --- Continued}
\end{table*}

\clearpage

\begin{table*}
 \begin{tabular}{c c c c c c c c c c c c}
  \hline
  \hline
No. & R.A. & DEC. &  $L_{\rm maj}$ & $L_{\rm min}$ & $\theta$ & $R_{\rm eff}$  & ${T}_{\rm dust}$& 
n(H$_{2}$) & $M$ & $M_{\rm BE}$ & Type\\
 & (J2000) & (J2000) & (${}''$) & (${}''$)& ($^{\circ}$)& (pc)  &(K)  & ($\rm 10^{5}\ cm^{-3}$) & ($\rm M_\odot$) & ($\rm M_\odot$) & \\
  \hline\noalign{\smallskip}
CMCHerschel$-$161  &  04:28:54.732  &  +36:24:26.83  &    50.6  &    25.1  &    68.4  &    0.03  &  14.8  &     0.8  &     0.7  &     0.6  &  PRE  \\
CMCHerschel$-$162  &  04:34:51.502  &  +36:24:24.01  &    59.2  &    36.5  &   138.3  &    0.05  &  13.9  &     0.4  &     1.3  &     0.9  &  PRE  \\
CMCHerschel$-$163  &  04:34:53.546  &  +36:23:21.73  &    60.6  &    37.6  &   117.3  &    0.05  &  13.2  &     1.4  &     4.4  &     0.9  &  PRO  \\
CMCHerschel$-$164  &  04:35:32.941  &  +36:17:15.68  &   124.6  &    20.2  &   153.4  &    0.04  &  14.3  &     0.3  &     0.3  &     0.7  &  USL  \\
CMCHerschel$-$165  &  04:35:38.926  &  +36:16:44.29  &    85.0  &    28.6  &    52.4  &    0.05  &  13.7  &     0.2  &     0.7  &     0.9  &  USL  \\
CMCHerschel$-$166  &  04:35:38.042  &  +36:15:40.07  &    81.5  &    44.5  &   135.4  &    0.06  &  13.4  &     0.2  &     1.7  &     1.2  &  PRE  \\
\hline
\end{tabular}
\centerline{Table \ref{Tab:herschelcs} --- Continued}
\end{table*}

\clearpage

\begin{table*}
 \begin{tabular}{c c c c c c c c c c c c}
  \hline
  \hline
No. & R.A. & DEC. &  $L_{\rm maj}$ & $L_{\rm min}$ & $\theta$ & $R_{\rm eff}$  & ${T}_{\rm dust}$& 
n(H$_{2}$) & $M$ & $M_{\rm BE}$ & Type\\
 & (J2000) & (J2000) & (${}''$) & (${}''$)& ($^{\circ}$)& (pc)  &(K)  & ($\rm 10^{5}\ cm^{-3}$) & ($\rm M_\odot$) & ($\rm M_\odot$) & \\
  \hline\noalign{\smallskip}
CMCHerschel$-$216  &  04:30:46.644  &  +35:44:14.13  &    72.7  &    27.4  &   132.7  &    0.04  &  13.9  &     0.7  &     1.4  &     0.8  &  PRE  \\
CMCHerschel$-$217  &  04:30:30.168  &  +35:43:56.97  &    78.8  &    35.5  &   123.6  &    0.05  &  13.0  &     0.6  &     2.5  &     1.0  &  PRE  \\
CMCHerschel$-$218  &  04:30:25.080  &  +35:42:38.41  &    56.1  &    26.0  &    48.6  &    0.03  &  14.5  &     0.6  &     0.7  &     0.7  &  PRE  \\
CMCHerschel$-$219  &  04:31:03.539  &  +35:42:20.52  &   171.0  &    24.4  &    46.7  &    0.06  &  14.2  &     0.3  &     1.8  &     1.1  &  PRE  \\
CMCHerschel$-$220  &  04:31:24.386  &  +35:39:51.01  &    84.6  &    23.4  &    19.8  &    0.04  &  13.3  &     0.9  &     1.5  &     0.8  &  PRE  \\
CMCHerschel$-$221  &  04:31:29.431  &  +35:39:30.31  &    58.5  &    23.0  &   149.8  &    0.03  &  13.5  &     1.1  &     0.9  &     0.6  &  PRE  \\
\hline
\end{tabular}
\centerline{Table \ref{Tab:herschelcs} --- Continued}
\end{table*}

\clearpage

\begin{table*}
 \begin{tabular}{c c c c c c c c c c c c}
  \hline
  \hline
No. & R.A. & DEC. &  $L_{\rm maj}$ & $L_{\rm min}$ & $\theta$ & $R_{\rm eff}$  & ${T}_{\rm dust}$
& n(H$_{2}$) & $M$ & $M_{\rm BE}$ & Type\\
 & (J2000) & (J2000) & (${}''$) & (${}''$)& ($^{\circ}$)& (pc)  &(K)  & ($\rm 10^{5}\ cm^{-3}$) & ($\rm M_\odot$) & ($\rm M_\odot$) & \\
  \hline\noalign{\smallskip}
CMCHerschel$-$271  &  04:30:15.573  &  +35:12:04.36  &    62.7  &    22.9  &   161.6  &    0.03  &  16.5  &     4.4  &     4.0  &     0.6  &  PRE  \\
CMCHerschel$-$272  &  04:30:52.976  &  +35:11:35.49  &    28.9  &    19.4  &    78.6  &    0.01  &  15.9  &     3.2  &     0.2  &     0.3  &  USL  \\
CMCHerschel$-$273  &  04:30:11.902  &  +35:10:57.55  &    69.1  &    25.2  &   176.6  &    0.04  &  16.4  &     1.1  &     1.7  &     0.7  &  PRE  \\
CMCHerschel$-$274  &  04:30:26.045  &  +35:10:05.81  &    83.9  &    21.9  &     8.6  &    0.03  &  15.5  &     3.4  &     4.0  &     0.7  &  PRE  \\
CMCHerschel$-$275  &  04:30:20.493  &  +35:09:25.78  &    87.0  &    26.6  &   152.6  &    0.04  &  14.8  &     1.4  &     3.6  &     0.9  &  PRE  \\
CMCHerschel$-$276  &  04:30:28.654  &  +35:09:11.42  &    57.5  &    21.3  &   126.6  &    0.03  &  15.7  &     8.8  &     4.9  &     0.5  &  PRO  \\
\hline
\end{tabular}
\centerline{Table \ref{Tab:herschelcs} --- Continued}
\end{table*}

\begin{table}
%\scriptsize
\tiny
 \setlength{\tabcolsep}{1.5pt}
\caption{Properties of molecular lines.}
\label{Tab:mlines}
\begin{tabular}{c c c c c c c c}
\hline
\noalign{\smallskip}
\multicolumn{7}{c}{$\rm H^{13}CO^{+}$ $J=1-0$}\\
\hline
\noalign{\smallskip}
No. &     $ T_{\rm mb}$  & $\int T_{\rm mb}d\upsilon $  & FWHM   & $V_{\rm LSR}$  & $\tau $ &$N$ & $X$\\
           &           (K)  &($\rm K\ km\ s^{-1}$)     &($\rm km\ s^{-1}$)  &($\rm km\ s^{-1}$) & &
$(\rm10^{12} cm^{-2})$        & ($\rm10^{-11}$)     \\           
\noalign{\smallskip}    
   \hline
  CMC$-$1&      1.18(0.05)&      0.87(0.05)&      0.69(0.49)&     -3.63(0.09)&    0.21&    2.47&   3.47\\
  CMC$-$2&            -&            -&            -&            -&       -&       -&      -\\
  CMC$-$3&      0.58(0.05)&      0.41(0.08)&      0.68(7.10)&     -2.40(0.03)&       -&       -&      -\\
  CMC$-$4&      0.45(0.05)&      0.33(0.05)&      0.67(0.81)&     -2.28(0.05)&       -&       -&      -\\
  CMC$-$5&      0.58(0.05)&      0.60(0.06)&      0.98(0.11)&     -1.23(0.06)&       -&       -&      -\\
\hline
\end{tabular}
%\ec
%
\tablefoot{\\
$T_{mb}$ is the corrected main-beam temperature.\\
$\int T_{mb}d\upsilon $ is the integrated main-beam temperature $T_{mb}$. \\
FWHM is the full width at half-maximum of the Gaussian fitting profile of the molecular line. \\
The FWHM of $\rm HCN (1-0)$ is the main component $\rm J F=(12-01)$. \\
The FWHM of $\rm N_{2}H^{+} (1-0)$ is its first component $\rm J F_{1} F=(101-012)$. \\
$V_{LSR}$ is the local standard of rest velocity.\\
$N$ is the column density.\\
$X$ is the abundance.\\
$N$ and $X$ for $\rm HCN$ and $\rm N_{2}H^{+}$ are derived from their component $\rm J F=(12-01)$ 
and $\rm J F_{1} F=(101-012),$ respectively.\\
Optical depth $\tau$ for $\rm C_{2}H$ is estimated by its main component $\rm JF=(3/2,2-1/2,1)$ 
with the HFS method in the CLASS software.\\
Optical depth $\tau$ for $\rm HCN$ is estimated by its main component $\rm JF=(12-01)$. \\
Optical depth $\tau$ for $\rm N_{2}H^{+}$ is estimated by its first component $\rm JF_{1} F=(101-012)$ 
}
\end{table}

\clearpage

\begin{table}
\tiny
 \setlength{\tabcolsep}{1.5pt}
\begin{tabular}{c c c c c c c c}
\hline
\noalign{\smallskip}
\multicolumn{7}{c}{$\rm HN^{13}C$ $J=1-0$}\\
\hline
\noalign{\smallskip}
No. &     $ T_{\rm mb}$  & $\int T_{\rm mb}d\upsilon $  & FWHM    & $V_{\rm LSR}$   & $\tau $     & 
$N$ & $X$\\
           &           (K)  &($\rm K\ km\ s^{-1}$)     &($\rm km\ s^{-1}$)  &($\rm km\ s^{-1}$) & &
$(\rm 10^{12}cm^{-2})$        & ($\rm10^{-11}$)     \\           
\noalign{\smallskip}    

   \hline
  CMC$-$1&      0.45(0.06)&      0.43(0.07)&      0.90(0.17)&     -3.52(0.09)&    0.11&    1.34&   1.88\\
  CMC$-$2&            -&            -&            -&            -&       -&       -&      -\\
  CMC$-$3&      0.22(0.05)&      0.29(0.08)&      1.24(0.51)&     -2.27(0.17)&    0.38&    0.86&   1.88\\
  CMC$-$4&      0.17(0.05)&      0.32(0.11)&      1.72(0.89)&     -1.92(0.23)&    0.42&    0.57&   1.15\\
  CMC$-$5&            -&            -&            -&            -&       -&       -&      -\\
\hline
\end{tabular}
\centerline{Table \ref{Tab:mlines} --- Continued}

\end{table}

\begin{table}
\tiny
 \setlength{\tabcolsep}{1.5pt}
\begin{tabular}{c c c c c c c c}
\hline
\noalign{\smallskip}
\multicolumn{7}{c}{$\rm C_{2}H$ $J=1-0$}\\
\hline
\noalign{\smallskip}
No. &     $ T_{\rm mb}$  & $\int T_{\rm mb}d\upsilon $  & FWHM  & $V_{\rm LSR}$ &$\tau $        & 
$N$ & $X$\\
           &           (K)  &($\rm K\ km\ s^{-1}$)     &($\rm km\ s^{-1}$)  &($\rm km\ s^{-1}$) & &
$(\rm 10^{14}cm^{-2})$        & ($\rm10^{-9}$)     \\           
\noalign{\smallskip}    

   \hline
     CMC$-$1&      1.34(0.06)&      1.20(0.07)&      0.85(0.14)&     -3.65(0.04)&      0.10(0.12)&    0.47&   0.66\\
  CMC$-$2&            -&            -&            -&            -&            -&       -&      -\\
  CMC$-$3&      0.46(0.05)&      0.33(0.06)&      0.67(0.17)&     -2.36(0.09)&      6.74(3.35)&    0.71&   1.55\\
  CMC$-$4&      0.24(0.06)&      0.17(0.06)&      0.67(0.52)&     -2.37(0.17)&     15.84(6.06)&    0.86&   1.74\\
  CMC$-$5&      0.44(0.06)&      0.52(0.07)&      1.10(0.16)&     -1.19(0.09)&      1.81(1.58)&    0.36&   0.77\\
\hline
\end{tabular}
\centerline{Table \ref{Tab:mlines} --- Continued}

\end{table}

\begin{table}
\scriptsize
\setlength{\tabcolsep}{1.5pt}
\begin{tabular}{c c c c c c c c}
\hline
\noalign{\smallskip}
\multicolumn{8}{c}{HCN $J=1-0$}\\
\hline
\noalign{\smallskip}
No. &     $ T_{\rm mb}$  & $\int T_{\rm mb}d\upsilon $  & FWHM               & $V_{\rm LSR}$        &  $\tau $  & $N$         & $X$\\
           &           (K)  &($\rm K\ km\ s^{-1}$)     &($\rm km\ s^{-1}$)  &($\rm km\ s^{-1}$)& 
&$(\rm 10^{14}cm^{-2})$ & ($\rm 10^{-9}$)     \\           
\noalign{\smallskip}    

   \hline
  CMC$-$1&      2.96(0.05)&      3.32(0.07)&      1.05(0.02)&     -3.71(0.01)&      0.21(0.22)&    0.16&   0.23\\
  CMC$-$2&            -&            -&            -&            -&            -&       -&      -\\
  CMC$-$3&      0.19(0.05)&      0.39(0.09)&      1.89(0.52)&     -2.67(0.21)&     15.52(6.36)&    0.23&   0.49\\
  CMC$-$4&      0.24(0.05)&      0.35(0.06)&      1.38(0.23)&     -2.35(0.14)&      8.08(2.09)&    0.10&   0.21\\
  CMC$-$5&      0.34(0.05)&      0.82(0.09)&      2.24(0.27)&     -1.44(0.13)&     28.66(1.83)&    0.89&   1.87\\
\hline
\end{tabular}
\centerline{Table \ref{Tab:mlines} --- Continued}
\end{table}

\clearpage

\begin{table}
\tiny
 \setlength{\tabcolsep}{1.5pt}
\begin{tabular}{c c c c c c c c}
\hline
\noalign{\smallskip}
\multicolumn{7}{c}{${\rm HCO}^{+}$ $J=1-0$}\\
\hline
\noalign{\smallskip}
No. &     $ T_{\rm mb}$  & $\int T_{\rm mb}d\upsilon $  & FWHM              & $V_{\rm LSR}$ & $\tau $       & $N$ & $X$\\
           &           (K)  &($\rm K\ km\ s^{-1}$)     &($\rm km\ s^{-1}$)  &($\rm km\ s^{-1}$) & &
$(\rm 10^{14}cm^{-2})$        & ($\rm10^{-9}$)     \\           
\noalign{\smallskip}    
   \hline
  CMC$-$1&      6.16(0.07)&      8.25(0.10)&      1.26(0.02)&     -3.68(0.01)&   10.63&    1.24&   1.73\\
  CMC$-$2&            -&            -&            -&            -&       -&       -&      -\\
  CMC$-$3&            -&            -&            -&            -&       -&       -&      -\\
  CMC$-$4&            -&            -&            -&            -&       -&       -&      -\\
  CMC$-$5&      0.42(0.06)&      0.56(0.08)&      1.25(0.22)&      0.09(0.08)&       -&       -&      -\\
\hline
\end{tabular}
\centerline{Table \ref{Tab:mlines} --- Continued}

\end{table}

\begin{table}
\tiny
 \setlength{\tabcolsep}{1.5pt}
\begin{tabular}{c c c c c c c c}
\hline
\noalign{\smallskip}
\multicolumn{7}{c}{HNC $J=1-0$}\\
\hline
\noalign{\smallskip}
No. &     $ T_{\rm mb}$  & $\int T_{\rm mb}d\upsilon $  & FWHM              & $V_{\rm LSR}$  & $\tau $      & $N$ & $X$\\
           &           (K)  &($\rm K\ km\ s^{-1}$)     &($\rm km\ s^{-1}$)  &($\rm km\ s^{-1}$) & &
$(\rm 10^{14}cm^{-2})$        & ($\rm10^{-9}$)     \\           
\noalign{\smallskip}    
   \hline
  CMC$-$1&      4.48(0.04)&      5.70(0.06)&      1.20(0.01)&     -3.67(0.01)&    5.26&    0.67&   0.94\\
  CMC$-$2&            -&            -&            -&            -&       -&       -&      -\\
  CMC$-$3&      0.69(0.05)&      1.18(0.08)&      1.60(0.13)&     -2.44(0.05)&   19.20&    0.43&   0.94\\
  CMC$-$4&      0.50(0.06)&      0.72(0.09)&      1.36(0.21)&     -2.09(0.08)&   20.78&    0.28&   0.58\\
  CMC$-$5&      0.77(0.05)&      1.78(0.09)&      2.17(0.12)&     -1.06(0.05)&       -&       -&      -\\
\hline
\end{tabular}
\centerline{Table \ref{Tab:mlines} --- Continued}

\end{table}

\begin{table}
\scriptsize
\setlength{\tabcolsep}{1.5pt}
\begin{tabular}{c c c c c c c c}
\hline
\noalign{\smallskip}
\multicolumn{7}{c}{$\rm N_{2}H^{+}$ $J=1-0$}\\
\hline
\noalign{\smallskip}
No. &     $ T_{\rm mb}$  & $\int T_{\rm mb}d\upsilon $  & FWHM & $V_{\rm LSR}$  & $\tau $  & 
$N$ & $X$\\
           &           (K)  &($\rm K\ km\ s^{-1}$)     &($\rm km\ s^{-1}$)  &($\rm km\ s^{-1}$) & &
$(\rm 10^{13}cm^{-2})$        & ($\rm10^{-10}$)     \\           
\noalign{\smallskip}     

   \hline
  CMC$-$1&      1.41(0.06)&      1.05(0.07)&      0.70(0.05)&    -11.63(0.04)&      0.10(0.01)&    1.69&   2.38\\
  CMC$-$2&            -&            -&            -&            -&            -&       -&      -\\
  CMC$-$3&      0.41(0.05)&      0.38(0.06)&      0.85(0.16)&    -10.36(0.09)&      0.83(1.49)&    0.73&   1.60\\
  CMC$-$4&      0.38(0.06)&      0.26(0.06)&      0.63(0.29)&    -10.34(0.12)&      0.58(0.50)&    0.45&   0.91\\
  CMC$-$5&      0.30(0.06)&      0.20(0.04)&      0.63(0.14)&     -9.14(0.10)&      0.10(0.16)&    0.28&   0.59\\
\hline
\end{tabular}
\centerline{Table \ref{Tab:mlines} --- Continued}
\end{table}

\clearpage

\begin{table}
\tiny
 \setlength{\tabcolsep}{1.5pt}
\begin{tabular}{c c c c c c c }
\hline
\noalign{\smallskip}
\multicolumn{7}{c}{$\rm C^{18}O$ $J=1-0$}\\
\hline
\noalign{\smallskip}
No. &     $ T_{\rm mb}$  & $\int T_{\rm mb}d\upsilon $  & FWHM              & $V_{\rm LSR}$         & 
$N$ & $X$\\
           &           (K)  &($\rm K\ km\ s^{-1}$)     &($\rm km\ s^{-1}$)  &($\rm km\ s^{-1}$) & 
$(\rm 10^{15}cm^{-2})$        & ($\rm10^{-8}$)     \\           
\noalign{\smallskip}    

   \hline
  CMC$-$1&   0.96(0.08)&   1.21(0.09)&   1.18(0.11)&  -3.63(0.04)&    1.47&   2.06\\
  CMC$-$2&         -&         -&         -&         -&       -&      -\\
  CMC$-$3&   2.93(0.09)&   1.95(0.09)&   0.63(0.03)&  -2.33(0.02)&    2.06&   4.50\\
  CMC$-$4&   3.45(0.09)&   2.43(0.08)&   0.66(0.03)&  -2.24(0.02)&    2.56&   5.20\\
  CMC$-$5&   3.95(0.09)&   4.23(0.10)&   1.00(0.03)&  -1.21(0.01)&    4.52&   9.55\\
\hline
\end{tabular}
\centerline{Table \ref{Tab:mlines} --- Continued}

\end{table}

\begin{table}
\tiny
 \setlength{\tabcolsep}{1.5pt}
\begin{tabular}{c c c c c c c }
\hline
\noalign{\smallskip}
\multicolumn{7}{c}{$\rm ^{13}CO$ $J=1-0$}\\
\hline
\noalign{\smallskip}
No. &     $ T_{\rm mb}$  & $\int T_{\rm mb}d\upsilon $  & FWHM              & $V_{\rm LSR}$         & 
$N$ & $X$\\
           &           (K)  &($\rm K\ km\ s^{-1}$)     &($\rm km\ s^{-1}$)  &($\rm km\ s^{-1}$) & 
$(\rm 10^{16}cm^{-2})$        & ($\rm10^{-6}$)     \\           
\noalign{\smallskip}    

   \hline
  CMC$-$1&   6.40(0.09)&  14.31(0.15)&   2.10(0.02)&  -3.62(0.01)&    1.73&   0.24\\
  CMC$-$2&   0.72(0.09)&   2.47(0.17)&   3.24(0.22)&  -1.64(0.12)&    0.28&   0.11\\
  CMC$-$3&   2.66(0.10)&   4.85(0.14)&   1.71(0.05)&  -2.40(0.02)&    0.51&   0.11\\
  CMC$-$4&   2.84(0.09)&   5.13(0.12)&   1.70(0.05)&  -2.45(0.02)&    0.54&   0.11\\
  CMC$-$5&   4.00(0.09)&   9.00(0.14)&   2.12(0.04)&  -0.87(0.02)&    0.96&   0.20\\
\hline
\end{tabular}
\centerline{Table \ref{Tab:mlines} --- Continued}

\end{table}

\clearpage

%%%%%%%%%%%%%%%%%%%%%%%%%

\begin{table*}
\caption[]{Detection rates of the observed molecular lines ($J=1-0$) for 
14 protostellar cores, 
7 prestellar cores, and 
8 observation reference positions that are offset from the cores. }
\label{Tab:detecrate}
  \setlength{\tabcolsep}{3.0pt}
 \small
 \begin{tabular}{cccccccccc}

  \hline
  \hline
      No.
    & $\rm H^{13}CO^{+}$
    & $\rm HN^{13}C$ 
    & $\rm C_{2}H$
    & $\rm HCN$ 
    & $\rm {HCO}^{+}$ 
    & $\rm HNC$ 
    & $\rm N_{2}H^{+}$ 
    & $\rm C^{18}O$ 
    & $\rm ^{13}CO$ \\
  \hline
Protostellar
&79\%
&64\%
&100\%
&100\%
&100\%
&100\%
&71\%
&86\%
&100\%\\
Prestellar 
&86\%
&71\%
&71\%
&71\%
&57\%
&71\%
&71\%
&100\%
&100\%\\
Reference 
&50\%
&13\%
&88\%
&88\%
&88\%
&100\%
&38\%
&100\%
&100\%\\
\hline
\end{tabular}
\end{table*}
     
 \begin{table}
\caption[]{Parameters used to calculate the molecular column densities.}
\label{Tab:Paracalcu}
\setlength{\tabcolsep}{2pt}
\begin{tabular}{r l l l l l  }
  \hline
  \hline
Line Name        & $\nu $    & $E_{\rm u}/k$ &  $\mu$     &   $B_{\rm 0}$ &   $R_{i}$            \\
                 & (MHz)     & (K)         &   (D)      &   (MHz)       &                      \\
  \hline\noalign{\smallskip}
 $\rm H^{13}CO^{+}\ (1-0)$  &      86754.288  &        4.16  &  3.90     &  43377.3   &  1   \\
     $\rm HN^{13}C\ (1-0)$  &      87090.851  &        4.18  &  2.699    &  43545.61  &  1   \\
     $\rm C_{2}H\ (1-0)$  &      87316.898  &        4.19  &  0.769    &  43674.52  &  $\rm 0.416^{a}$\\
        $\rm HCN\ (1-0)$  &      88631.847  &        4.25  &  2.984    &  44315.97  &  $\rm 5/9^{b}$ \\
   $\rm {HCO}^{+}\ (1-0)$ &      89188.526  &        4.28  &  3.888    &  44594.4   &  1  \\
        $\rm HNC\ (1-0)$  &      90663.568  &        4.35  &  3.05     &  45331.98  &  1  \\
 $\rm N_{2}H^{+}\ (1-0)$  &      93173.777  &        4.47  &  3.40     &  46586.87  &  $\rm 1/9^{c}$ \\
      $\rm C^{18}O\ (1-0)$  &     109782.176  &        5.27  &  0.11049  &  54891.42  &  1   \\
      $\rm ^{13}CO\ (1-0)$  &     110201.3543  &       5.29  &  0.11046  &  55101.01 &  1   \\
\hline
\end{tabular}
\tablefoot{\\
$\nu $ is the rest frequency.\\
$E_{\rm u}$ is the upper energy. k is the Boltzmann constant.\\
$\mu$ is the permanent dipole moment along the axis of symmetry.\\
$B_{\rm 0}$ is the rigid rotor rotation constant.\\
$R_{i}$ is the relative intensity. \\
a: The relative intensity for $\rm C_{2}H\ JF=(3/2,2-1/2,1)$. \\
b: The relative intensity for $\rm N_{2}H^{+}\ JF_{1} F=(101-012)$.\\
c: The relative intensity for $\rm HCN\ JF=(12-01)$.
}
\end{table}

\begin{table*}
 \caption[]{Statistics of abundance ratios of 14 protostellar and 7 prestellar cores}
 \label{Tab:statabuncores} 
 \small
 \begin{tabular}{c c c c c c c c c c}
  \hline
  \hline
& $\rm H^{13}CO^{+}$ 
& $\rm HN^{13}C$     
& $\rm C_{2}H$       
& $\rm HCN$          
& $\rm {HCO}^{+}$      
& $\rm HNC$  
& $\rm N_{2}H^{+}$ 
& $\rm C^{18}O$
& $\rm ^{13}CO$\\
  & ($\rm{10}^{-11}$) 
& ($\rm {10}^{-11}$) 
& ($\rm {10}^{-9}$) 
& ($\rm {10}^{-9}$)
& ($\rm {10}^{-9}$)
& ($\rm {10}^{-9}$)
& ($\rm {10}^{-10}$)
& ($\rm {10}^{-8}$)
& ($\rm {10}^{-6}$)\\
\hline
Max.&      6.62  &      6.77  &      8.18  &      4.80  &      3.31  &      3.39  &      5.58  &     13.56  &      2.55  \\
Min.&      0.76  &      1.23  &      0.36  &      0.08  &      0.38  &      0.62  &      0.52  &      1.20  &      0.09  \\
Mean&      3.14  &      2.38  &      1.99  &      1.00  &      1.57  &      1.19  &      2.04  &      6.37  &      0.46  \\
Median&      2.69  &      1.97  &      1.55  &      0.40  &      1.35  &      0.99  &      1.67  &      5.99  &      0.21  \\
Std.&      1.98  &      1.49  &      1.92  &      1.35  &      0.99  &      0.75  &      1.52  &      3.67  &      0.60  \\
\hline
\end{tabular}
\end{table*}

\clearpage

\begin{table*}
\caption{Molecular line skewed profile.}
\label{Tab:Asymmetry}
 \begin{tabular}{c c c c c c c c}
  \hline
  \hline
     No. & ${V}_{\rm thick}$    & ${V}_{\rm thick}$& ${V}_{\rm thin}$         & $\Delta V$      & $\delta V$           & $\delta V$  &  Profile \\
          & $\rm {HCO}^{+} (1-0)$ & $\rm HNC (1-0)$   & $\rm {C}^{18}{O} (1-0)$ &$\rm {C}^{18}{O} (1-0)$ & 
          $\rm  {HCO}^{+} (1-0)$             &$\rm  HNC (1-0)$   &          \\
          & $\rm (km\ s^{-1})$             & $\rm (km\ s^{-1})$         &   $\rm (km\ s^{-1})$                &  $\rm (km\ s^{-1})$           &                     &             &           \\
  \hline
   CMC$-$1  &  -3.68(0.01)  &  -3.67(-3.67)  &  -3.63(0.04)  &  1.18(0.11)  &  -0.05(0.05)  &  -0.04(0.05)  &  N,N\\
   CMC$-$2  &  -  &  -  &  -  &  -  &  -  &  -  &  -,-\\
   CMC$-$3  &  -  &  -2.44(-2.44)  &  -2.33(0.02)  &  0.63(0.03)  &  -  &  -0.18(0.12)  &  -,N\\
   CMC$-$4  &  -  &  -2.09(-2.09)  &  -2.24(0.02)  &  0.66(0.03)  &  -  &  0.23(0.16)  &  -,N\\
   CMC$-$5  &  0.09(0.08)  &  -1.06(-1.06)  &  -1.21(0.01)  &  1.00(0.03)  &  1.30(0.13)  &  0.15(0.07)  &  R,N\\
   CMC$-$6  &  -0.09(0.02)  &  -0.43(-0.43)  &  -0.46(0.02)  &  1.11(0.05)  &  0.33(0.05)  &  0.02(0.04)  &  R,N\\
   CMC$-$7  &  -  &  -1.87(-1.87)  &  -1.77(0.02)  &  0.73(0.07)  &  -  &  -0.14(0.08)  &  -,N\\
   CMC$-$8  &  -2.94(0.03)  &  -2.85(-2.85)  &  -2.34(0.02)  &  1.02(0.05)  &  -0.58(0.08)  &  -0.50(0.06)  &  B,B\\
   CMC$-$9  &  -2.52(0.08)  &  -1.80(-1.80)  &  -1.83(0.01)  &  0.78(0.03)  &  -0.88(0.17)  &  0.04(0.08)  &  B,N\\
  CMC$-$10  &  -  &  -  &  -1.81(0.01)  &  0.76(0.05)  &  -  &  -  &  -,-\\
  CMC$-$11  &  -2.05(0.05)  &  -  &  -1.42(0.01)  &  0.68(0.02)  &  -0.92(0.11)  &  -  &  B,-\\
  CMC$-$12  &  -1.67(0.04)  &  -1.48(-1.48)  &  -1.47(0.02)  &  0.71(0.03)  &  -0.27(0.09)  &  -0.01(0.06)  &  B,N\\
  CMC$-$13  &  -0.58(0.13)  &  -0.19(-0.19)  &  -0.13(0.01)  &  0.53(0.06)  &  -0.84(0.38)  &  -0.11(0.21)  &  B,N\\
  CMC$-$14  &  -1.08(0.04)  &  -0.98(-0.98)  &  -0.65(0.03)  &  0.88(0.06)  &  -0.49(0.11)  &  -0.37(0.10)  &  B,B\\
  CMC$-$15  &  -1.17(0.03)  &  -1.04(-1.04)  &  -1.21(0.02)  &  0.83(0.07)  &  0.04(0.07)  &  0.20(0.08)  &  N,N\\
  CMC$-$16  &  -1.16(0.02)  &  -1.19(-1.19)  &  -1.49(0.02)  &  1.07(0.04)  &  0.31(0.04)  &  0.28(0.04)  &  R,R\\
  CMC$-$17  &  -1.87(0.02)  &  -1.50(-1.50)  &  -1.10(0.03)  &  0.98(0.07)  &  -0.78(0.10)  &  -0.40(0.07)  &  B,B\\
  CMC$-$18  &  -1.84(0.02)  &  -1.88(-1.88)  &  -1.27(0.03)  &  1.57(0.07)  &  -0.36(0.05)  &  -0.39(0.05)  &  B,B\\
  CMC$-$19  &  -0.96(0.04)  &  -0.79(-0.79)  &  -1.01(0.01)  &  0.87(0.02)  &  0.05(0.05)  &  0.25(0.07)  &  N,R\\
  CMC$-$20  &  -0.49(0.01)  &  -0.54(-0.54)  &  -0.53(0.01)  &  0.93(0.02)  &  0.05(0.02)  &  -0.00(0.03)  &  N,N\\
  CMC$-$21  &  -0.50(0.03)  &  -0.52(-0.52)  &  -0.61(0.01)  &  0.98(0.02)  &  0.10(0.05)  &  0.09(0.04)  &  N,N\\
  CMC$-$22  &  -1.96(0.03)  &  -1.91(-1.91)  &  -  &  -  &  -  &  -  &  -,-\\
  CMC$-$23  &  -1.44(0.03)  &  -1.13(-1.13)  &  -  &  -  &  -  &  -  &  -,-\\
  CMC$-$24  &  0.07(0.04)  &  0.01(0.01)  &  -0.59(0.04)  &  0.73(0.06)  &  0.91(0.18)  &  0.84(0.33)  &  R,R\\
  CMC$-$25  &  -0.46(0.02)  &  -0.61(-0.61)  &  -0.85(0.01)  &  0.72(0.05)  &  0.54(0.09)  &  0.33(0.07)  &  R,R\\
  CMC$-$26  &  0.37(0.02)  &  0.10(0.10)  &  0.04(0.04)  &  1.52(0.10)  &  0.22(0.06)  &  0.04(0.05)  &  N,N\\
  CMC$-$27  &  0.30(0.03)  &  0.13(0.13)  &  0.46(0.07)  &  1.31(0.17)  &  -0.12(0.09)  &  -0.26(0.12)  &  N,B\\
  CMC$-$28  &  0.84(0.01)  &  0.74(0.74)  &  0.52(0.03)  &  1.38(0.08)  &  0.23(0.05)  &  0.16(0.05)  &  N,N\\
  CMC$-$29  &  -0.51(0.02)  &  -0.54(-0.54)  &  -0.70(0.03)  &  0.80(0.07)  &  0.24(0.08)  &  0.21(0.07)  &  N,N\\
  CMC$-$30  &  -0.47(0.04)  &  -0.41(-0.41)  &  -0.53(0.07)  &  0.53(0.27)  &  0.11(0.26)  &  0.23(0.27)  &  N,N\\
\hline
\end{tabular}
\tablefoot{The $\rm {HCO}^{+} (1-0)$ and $\rm HNC (1-0)$  profiles are evaluated as follows: B denotes a blue profile. R denotes a red profile. N denotes neither blue nor red.}

\end{table*}
   
\clearpage

   \begin{figure*}
   \centering
   \includegraphics[width=\hsize]{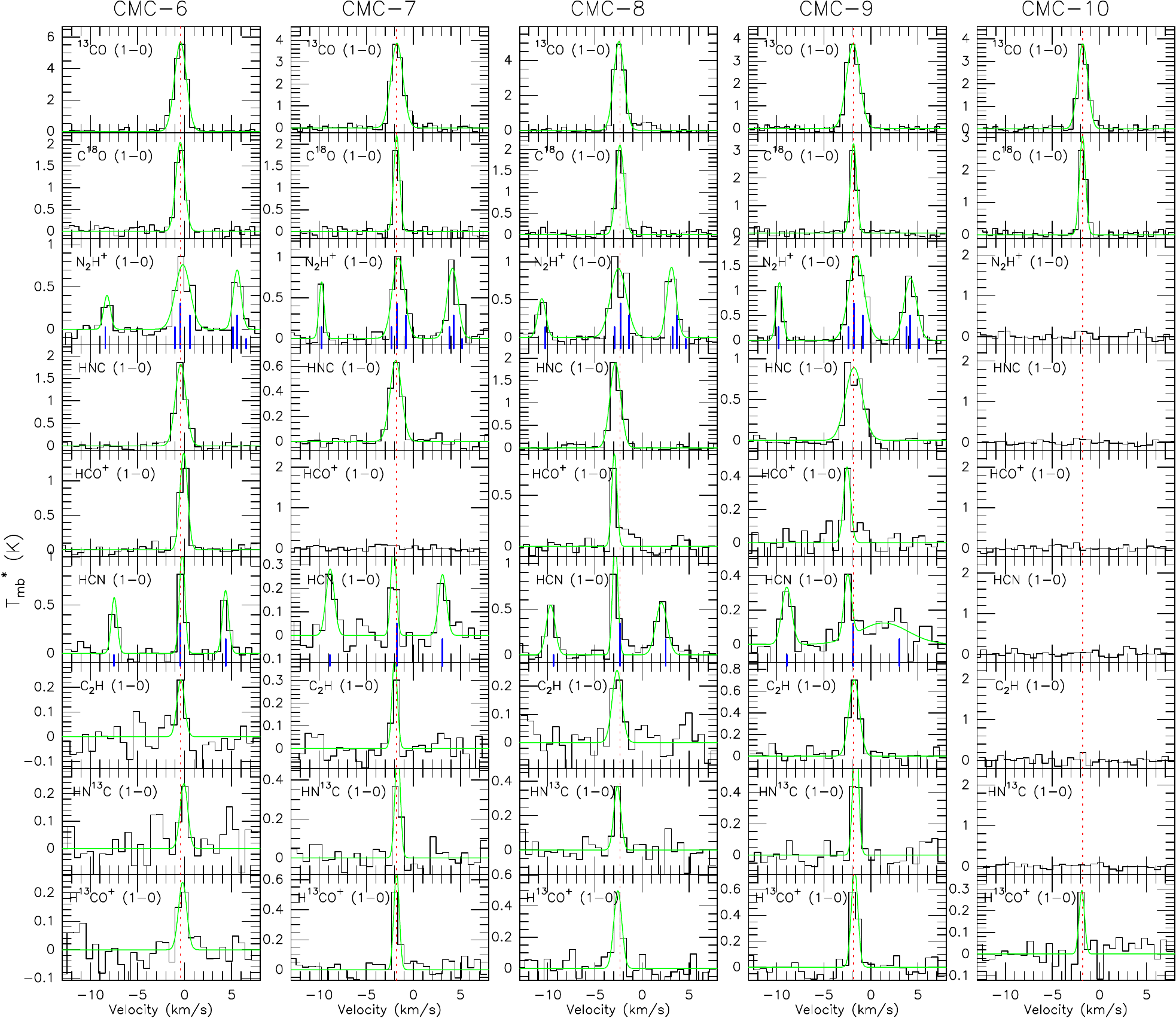}
   \centerline{Fig. \ref{Fig:cmclines} --- Continued}
   \end{figure*}
   
\clearpage

   \begin{figure*}
   \centering
   \includegraphics[width=\hsize]{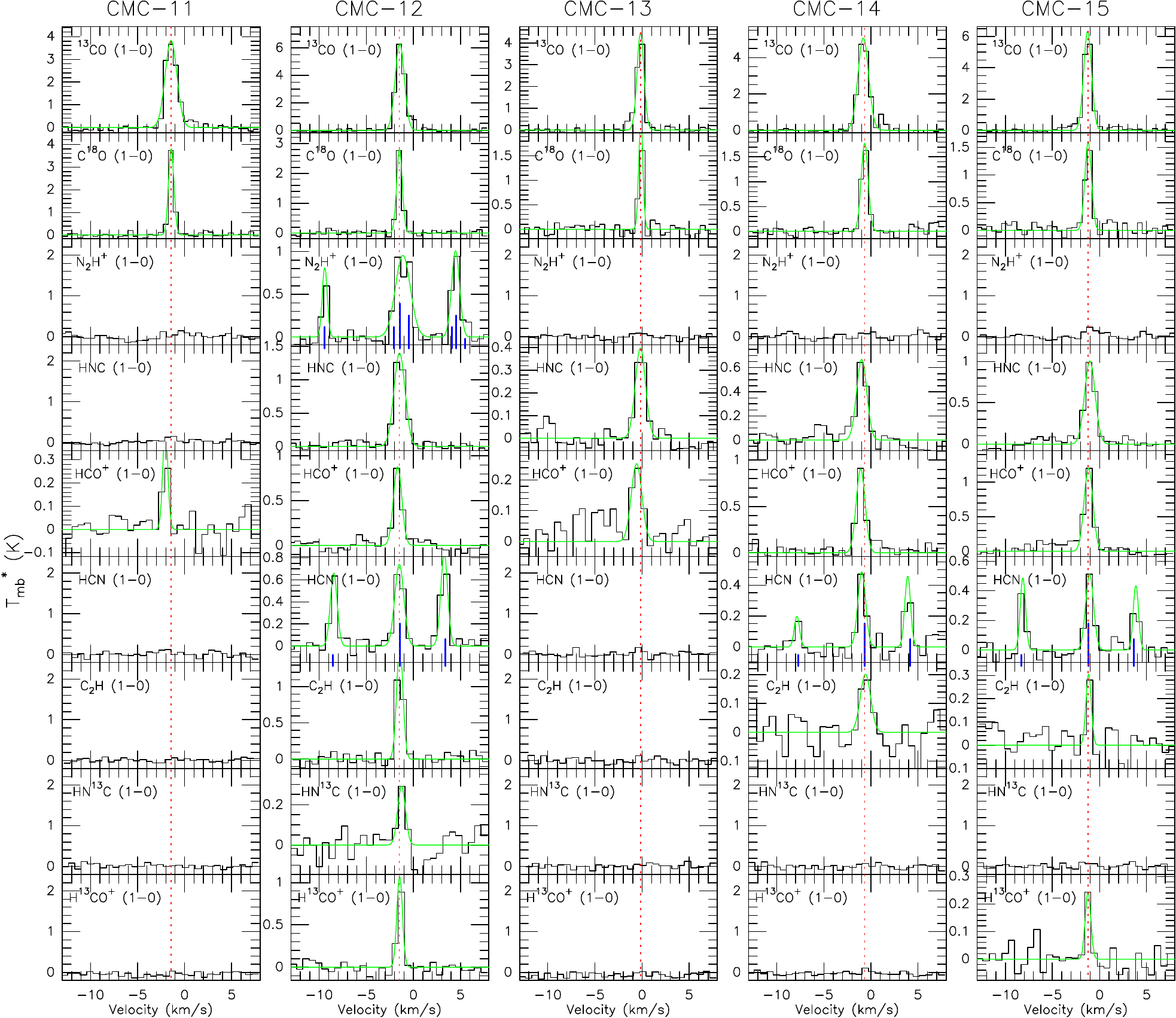}
   \centerline{Fig. \ref{Fig:cmclines} --- Continued}
   \end{figure*}
   
   \begin{figure*}
   \centering
   \includegraphics[width=\hsize]{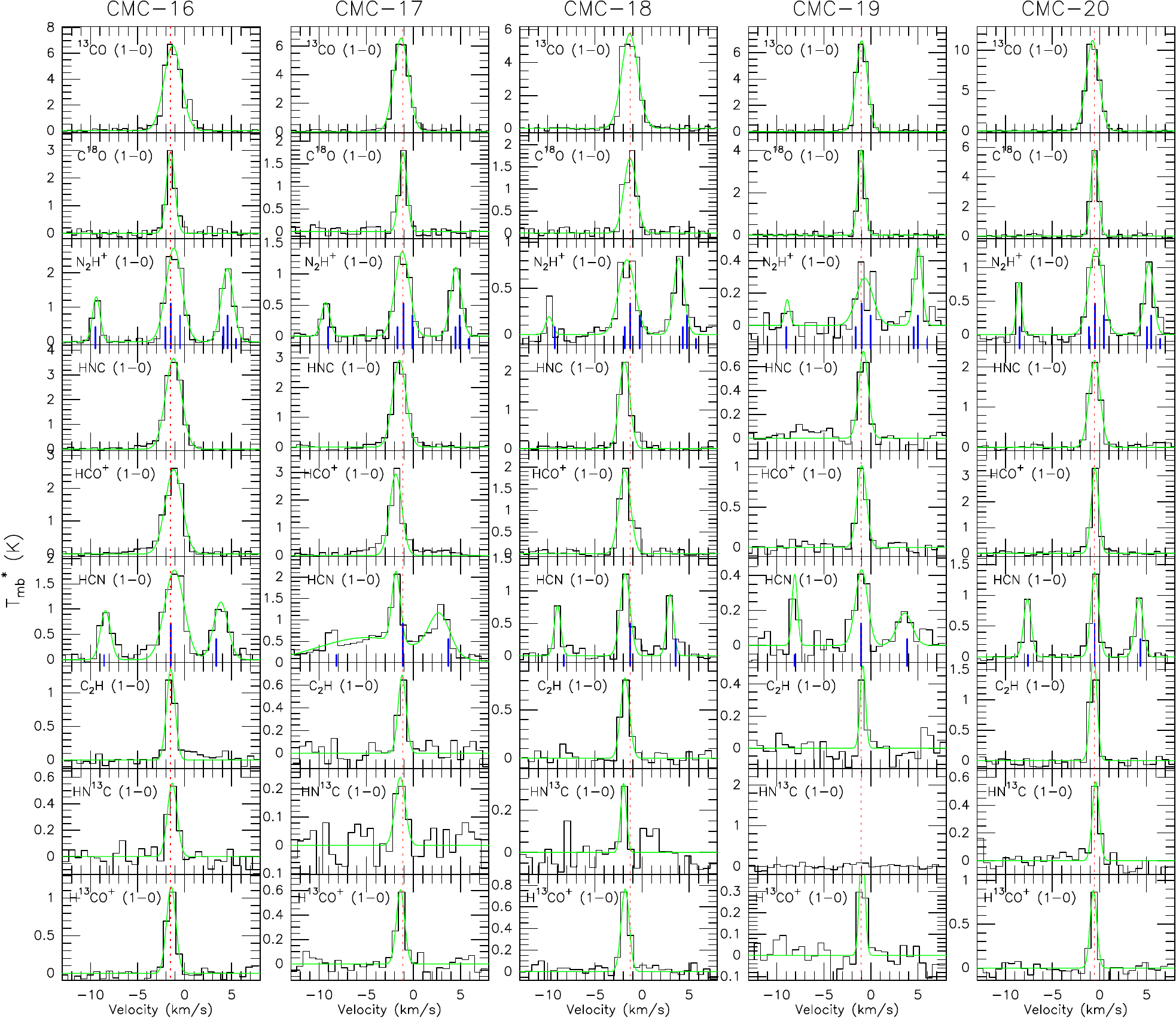}
   \centerline{Fig. \ref{Fig:cmclines} --- Continued}
   \end{figure*}
   
   \begin{figure*}
   \centering
   \includegraphics[width=\hsize]{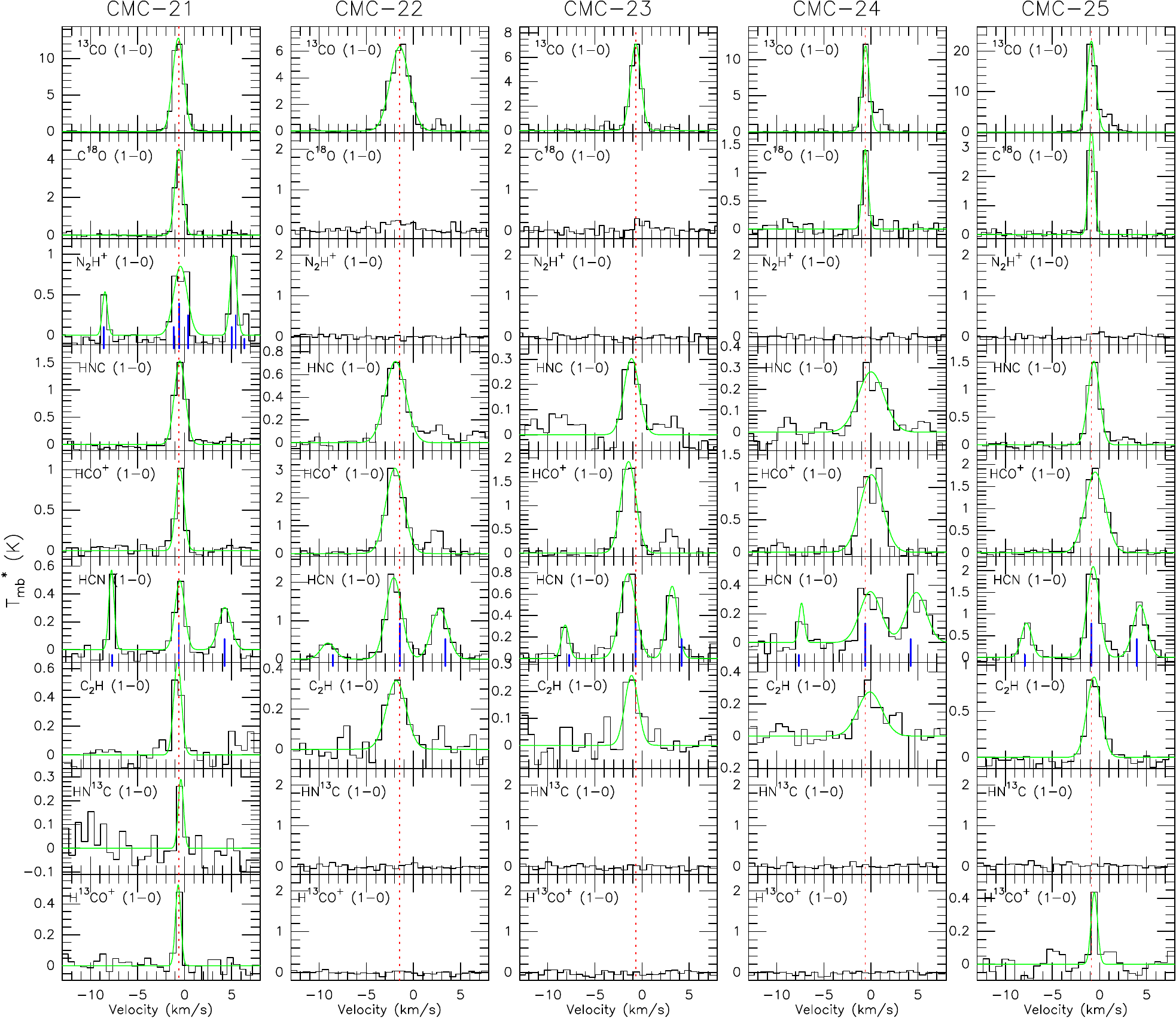}
   \centerline{Fig. \ref{Fig:cmclines} --- Continued}
   \end{figure*}
   
   \begin{figure*}
   \centering
   \includegraphics[width=\hsize]{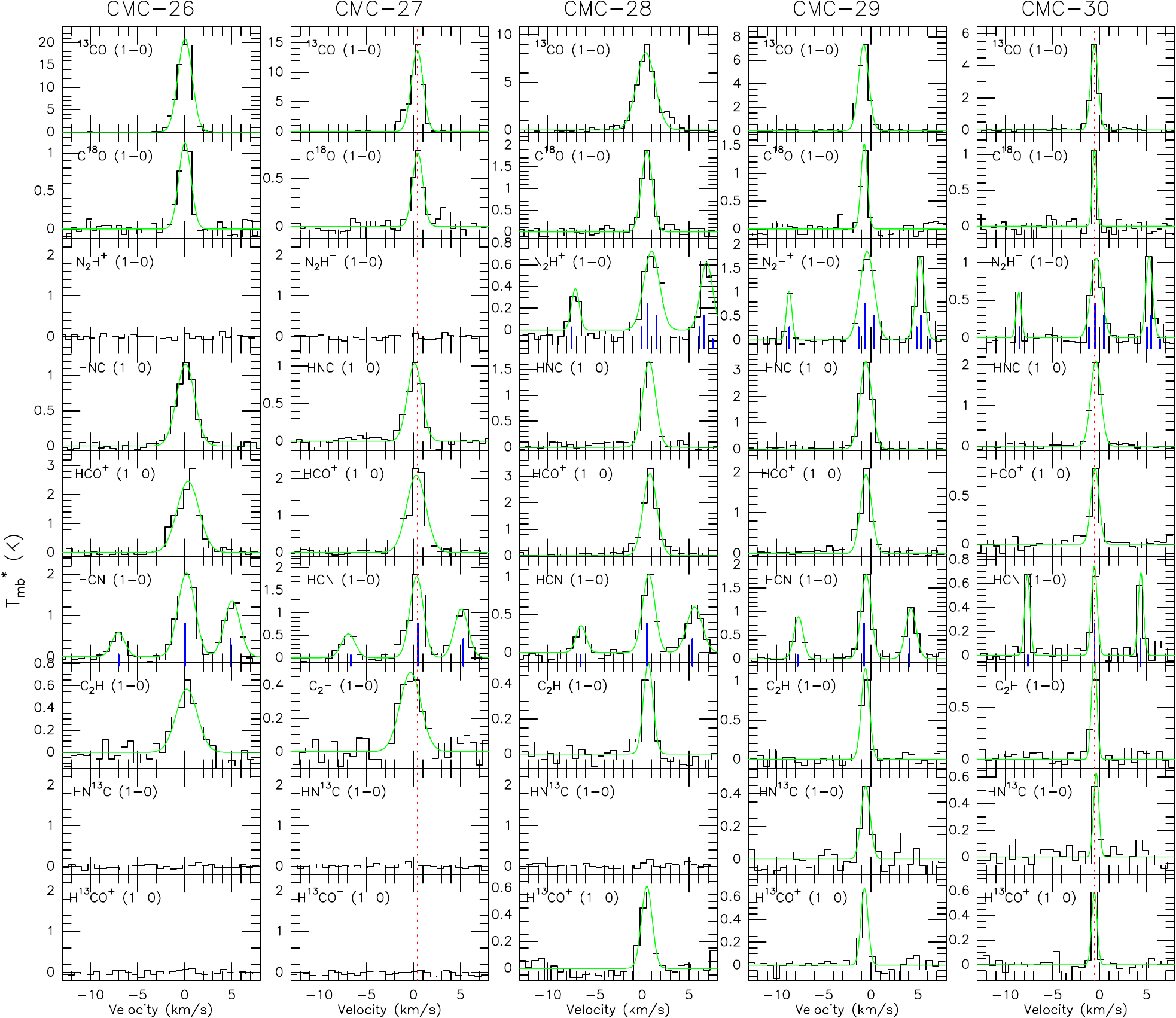}
   \centerline{Fig. \ref{Fig:cmclines} --- Continued}
   \end{figure*}
   
\clearpage

\end{document}